\documentclass[accepted=2020-10-18,10pt,letterpaper]{quantumarticle}
\pdfoutput=1
\usepackage[sort&compress,numbers]{natbib}
\usepackage[utf8]{inputenc}
\usepackage[english]{babel}
\usepackage[T1]{fontenc}
\usepackage{amsmath}
\usepackage{amsfonts}
\usepackage{amssymb}

\usepackage{bm}
\usepackage[pdftex]{graphicx}
\usepackage{enumerate}
\usepackage{bbold}
\usepackage{array}
\usepackage{multirow}
\usepackage{booktabs}
\usepackage[caption=false]{subfig}

\usepackage{hyperref}
\hypersetup{
    colorlinks=true,       
    linkcolor=cyan,          
    citecolor=magenta,        
    filecolor=magenta,      
    urlcolor=cyan,           
    runcolor=cyan
}

\begin{document}


\newcommand{\ket}[1]{\lvert#1\rangle}
\newcommand{\bra}[1]{\langle#1\rvert}
\newcommand{\be}{\begin{equation}}
\newcommand{\ee}{\end{equation}}
\newcommand{\ra}{\rangle}
\newcommand{\la}{\langle}

\newcommand{\red}[1]{\textcolor{red}{#1}}
\newcommand{\green}[1]{\textcolor{black}{#1}}
\newcommand{\blue}[1]{\textcolor{blue}{#1}}
\newcommand{\mg}[1]{\textcolor{magenta}{#1}}
\newcommand{\RM}[1]{\red{[RM:{#1}]}}

 
\title{Always-On Quantum Error Tracking with Continuous Parity Measurements}

\author{Razieh Mohseninia}
\affiliation{Institute for Quantum Studies, Chapman University, Orange, CA 92866, USA}
\affiliation{Departments of Electrical Engineering, University of Southern California, Los Angeles, California 90089, USA}

\author{Jing Yang}
\affiliation{Department of Physics and Astronomy, University of Rochester, Rochester, New York 14627, USA}

\author{Irfan Siddiqi}
\affiliation{Center for Quantum Coherent Science, Berkeley, CA 94720 USA}
\affiliation{Department of Physics, University of California, Berkeley, CA 94720 USA}

\author{Andrew N. Jordan}
\affiliation{Department of Physics and Astronomy, University of Rochester, Rochester, New York 14627, USA}
\affiliation{Institute for Quantum Studies, Chapman University, Orange, CA 92866, USA}
\orcid{0000-0002-9646-7013}

\author{Justin Dressel}
\affiliation{Institute for Quantum Studies, Chapman University, Orange, CA 92866, USA}
\affiliation{Schmid College of Science and Technology, Chapman University, Orange, CA 92866, USA}
\orcid{0000-0001-7216-1581}


\begin{abstract}
We investigate quantum error correction using continuous parity measurements to correct bit-flip errors with the three-qubit code. Continuous monitoring of errors brings the benefit of a continuous stream of information, which facilitates passive error tracking in real time. It reduces overhead from the standard gate-based approach that periodically entangles and measures additional ancilla qubits. However, the noisy analog signals from continuous parity measurements mandate more complicated signal processing to interpret syndromes accurately. We analyze the performance of several practical filtering methods for continuous error correction and demonstrate that they are viable alternatives to the standard ancilla-based approach. As an optimal filter, we discuss an unnormalized (linear) Bayesian filter, with improved computational efficiency compared to the related Wonham filter introduced by Mabuchi [New J. Phys. 11, 105044 (2009)]. We compare this optimal continuous filter to two practical variations of the simplest periodic boxcar-averaging-and-thresholding filter, targeting real-time hardware implementations with low-latency circuitry. As variations, we introduce a non-Markovian ``half-boxcar'' filter and a Markovian filter with a second adjustable threshold; these filters eliminate the dominant source of error in the boxcar filter, and compare favorably to the optimal filter. For each filter, we derive analytic results for the decay in average fidelity and verify them with numerical simulations. 
\end{abstract}

\maketitle

\section{Introduction}\label{sec:intro}
Quantum error correction (QEC) is essential to building a scalable and fault-tolerant quantum computer \cite{lidar2013quantum,gaitan2013quantum}. Although the theory of QEC has been developing since the 1990s and is now well established for the circuit model of quantum computation, the practical implemention of QEC in realistic hardware raises additional nuance that prompts more detailed investigation. The present work addresses one aspect of QEC implementation that is relevant to modern superconducting qubit architectures \cite{devoret2004superconducting,devoret2013superconducting} by investigating whether the time-continuous nature of standard dispersive qubit measurements can be used in principle to improve the logical-state-tracking fidelity in the prototypical error correction scenario of a 3-qubit bit-flip code \cite{Nielsen2011}. We show that direct monitoring of the error syndromes reduces hardware resources compared to the circuit model of QEC while maintaining performance.

Circuit models of QEC redundantly encode a logical qubit state into multiple physical qubits. Examples of this model include the Shor \cite{shor1995schemefor}, Steane \cite{steane1996error}, and Calderbank-Steane-Shor (CSS) codes \cite{calderbank1996good,steane1996simple}, as well as more general stabilizer codes \cite{gottesman1996class,gottesman1997stabilizer,gottesman1998theory,gottesman2009anintroduction,calderbank1998quantum}. Encoded information is checked by measuring ancillary qubits that are entangled with the redundant code subspaces. The ancilla measurements project the logical qubits back onto the code subspaces, effectively converting analog drifts of the encoded state into digital jumps between code subspaces (e.g., bit flips, phase flips, or combinations thereof). The measurement results provide information about jumps between code subspaces, thus enabling correct decoding of the logical qubit state. Different encoding schemes protect against different error types and quantities according to the redundancy of the code subspaces, with the simplest codes protecting against only single jumps per measurement cycle. Simple forms of such gate-based QEC have already been implemented in several experiments, see e.g.  \cite{reed2012realization,barends2014superconducting,chow2014implementing}. 

Stabilizer codes typically assume ancilla-based projective syndrome measurements. However, for superconducting qubits this assumption can be problematic for two reasons. First, the repeated entangling and disentangling of the code and ancillary qubits adds additional gate overhead and hardware resources. This overhead also increases the vulnerability of the protocol to additional error mechanisms. Second, superconducting qubit architectures implement projective measurements by integrating and thresholding time-continuous dispersive measurements, which are not instantaneous projections as assumed by the theoretical quantum circuit model \cite{murch2013observing,weber2014mapping,ficheux2017observing,campagne2016observing}. The temporally extended nature of the measurements further increases the overhead by substantially lengthening the achievable cycle time for periodic syndrome measurements. These challenges raise the question whether alternative strategies for performing the syndrome measurements could be fruitful. 

A possible route to perform QEC without the overhead of ancilla qubits is to directly monitor the error syndromes continuously in time \cite{Wiseman2009,jacobs2014quantum}. With this variation, the code subspaces for the error syndromes are directly coupled to a continuous readout device \cite{mao2004mesoscopic,trauzettel2006parity,williams2008entanglement,lalumiere2010tunable,tornberg2010high,haack2010parity,zu2014demand}, avoiding the need for periodic entangling gates and additional ancilla measurements. This idea of continuous quantum error correction was proposed in Ref.~\cite{ahn2002continuous}, and further developed in Ref.~\cite{ahn2003quantum,sarovar2004practical,vanhandel2005optimal,mabuchi2009continuous, denhez2012quantum,
hsu2016methodfor,atalaya2017bacon,Cardona2019}. Experiments have since demonstrated several necessary components of continuous QEC, including measurement-generated entanglement between pairs of qubits via continuous parity and partial-parity measurements in superconducting circuits \cite{riste2013deterministic,PhysRevLett.112.170501,PhysRevX.6.041052,PhysRevX.6.031036}. We have thus reached a stage of technological development where implementing continuous QEC becomes feasible for at least the simplest codes.

In principle, continuous measurements have the advantages of being:
(1) Always on - A continuous measurement eliminates dead time between measurement cycles of ancillary qubits, preventing errors from occurring during entangling-gate sequences.
(2) Natural - Standard dispersive measurements in superconducting circuitry are already continuous, producing binary results only after integrating and thresholding. 
(3) Potentially faster - Continuous measurements have a characteristic time scale to distinguish the signal from the intrinsic background noise, which can be shortened to yield ``strong continuous measurements'' that rapidly yield information about error syndromes. 
Continuous measurements also have disadvantages, however, since they are: (i) Noisy - An experimenter must interpret a stochastic time-continuous signal, which is a more difficult signal processing problem than for discrete ancilla measurements. (ii) Challenging - Using ancillary qubits of the same design as the data qubits is conceptually straightforward, whereas physically implementing direct syndrome measurements requires specialized qubit circuits. (iii) Computationally expensive - Optimal signal processing of the continuous readout may have high latency.   

In the present paper, we assess the performance of implementing continuous QEC for the simplest three-qubit bit-flip code, assuming a simplified model of modern superconducting hardware, and develop practical filters to interpret the stochastic time-continuous signals. 
We show that for {\it passive error tracking} the benefits of continuous measurements can outweigh the disadvantages, enabling high-fidelity decoding of the logical qubit without the need for active feedback. This positive result is particularly interesting, since much of the previous work on continuous QEC has focused on applying active feedback based on the monitored syndrome signals to also correct the errors continuously \cite{ahn2002continuous,Cardona2019}, which has been shown to be rather ineffective due to the large noise of the signal, as well as degradations from signal processing delay \cite{kumar2019quantum}. For simplicity, we consider passive error tracking for a prototypical setup that tracks only Poisson-distributed bit-flip errors in a three-qubit code, and consider possible generalizations in the subsequent discussion. However, we emphasize that these techniques also apply to active error correction with the additional caveat that additional errors not considered here may occur during the correction pulses.

We compare three signal-processing filters for interpreting the error syndromes. We expand upon the Bayesian filtering methods discussed by van Handel and Mabuchi \cite{vanhandel2005optimal,mabuchi2009continuous}
and derive a linear version of the Bayesian filter that permits faster numerical calculation of the most likely state compared to the nonlinear (Wonham) filter. To address the issue of computational expense we then propose two variations of the simplest Markovian ``boxcar'' filter that averages the noisy signals over temporal segments of a fixed length. After analyzing the ways in which error tracking can fail for the boxcar filter, we identify the dominant source of error that compromises its performance. We then introduce an improved non-Markovian ``half-boxcar'' filter that corrects the dominant error of the boxcar filter by re-examining the memory of the preceding half-boxcar average. Finally, we introduce an improved Markovian ``double threshold'' filter that also corrects the dominant error of the crude boxcar filter by using two signal thresholds. Both variations can be readily implemented with low-latency circuitry, such as field-programmable gate arrays (FPGAs), and compare favorably to the optimal Bayesian filter. We derive analytic results for the initial drop in fidelity and the approximately linear fidelity decay rate for each filter, optimize them over the free filter parameters, and verify them with numerical simulations, finding good agreement. 

We now summarize the main findings of this paper. We provide a simple direct parity readout implementation for superconducting transmons that relies on entangling strongly detuned resonator linewidths with the two-qubit parity subspaces. Our proposed design has recently been fabricated and tested \cite{LivingstonAPS2018,LivingstonAPS2019}, which motivates our current work. We derive analytic results for the initial drop of each signal-processing filter, as well as their logical error rates, which are summarized in Tables~\ref{tab:fingamma} and \ref{tab:optscaling}. We verify these analytical results through explicit numerical simulations in Figures~\ref{fig:fidelity}--\ref{fig:optimalgamma}. In particular, we conclude that the ``half-boxcar'' filter performs comparably to the optimal Bayesian filter and is therefore a good candidate for real-time laboratory implementation with an FPGA because of its simple numerical requirements. 

The paper is organized as follows. In Section~\ref{sec:code} we review the basics of the three-qubit bit-flip code, and pose the problem. In Section~\ref{sec:setup}, we discuss a possible implementation for the continuous syndrome measurements. In Section~\ref{sec:continuous} we introduce and analyze an optimal linear Bayesian filter. In Section~\ref{sec:periodic} we introduce and analyze three periodic averaging filters that are more efficient but suboptimal. In Section~\ref{sec:simulation}, we describe our numerical simulations for the continuous syndrome measurements. We verify our analytics of the continuous and periodic filters with the numerics, and discuss the results. We conclude in Section~\ref{sec:conclusions}. We also include an Appendix that contains a complementary analysis of an ancilla-based projective measurement implementation of the three-qubit bit-flip code. 


\section{Three-qubit bit-flip code}\label{sec:code}
For clarity we review the basics of the three-qubit bit-flip code and introduce notation and terminology. 

\subsection{Encoding and error syndromes}\label{sec:code:encoding}

The standard bit flip code redundantly encodes a logical qubit state $\alpha\ket{0}_{L}+\beta\ket{1}_{L}$
into three physical qubits,
\be \label{eq:encode0}
|\psi_0\ra = \alpha\ket{000}+\beta\ket{111},
\ee
and uses majority-voting to identify and correct single bit flip errors. We number the bits from left to right as 123. We use quantum computing conventions for the Pauli operators: $I = |0\ra\la0| + |1\ra\la1|$, $X = |0\ra\la1| + |1\ra\la0|$, $Y = -i|0\ra\la1| + i|1\ra\la0|$, and $Z = |0\ra\la0| - |1\ra\la1|$. To indicate idling and bit flip operations on the physical qubits, we use the Pauli identity $I$ and flip $X$ operators. The initial encoding of the logical state is recovered after an idle operation $III$. Omitting tensor products for brevity, we use the notation $III$ to indicate the original encoding. Similarly, the operations after a single bit flip on the first, second, or third qubit are $XII$, $IXI$, $IIX$, respectively, which also serve as suitable labels for the resulting encodings. For example, these bit flips produce the states
\begin{eqnarray}
|\psi_1\ra &=& XII|\psi_0\ra = \alpha |100\ra + \beta |011\ra, \\
|\psi_2\ra &=& IXI|\psi_0\ra = \alpha |010\ra + \beta |101\ra, \\
|\psi_3\ra &=& IIX|\psi_0\ra = \alpha |001\ra + \beta |110\ra.\label{eq:encode4}
\end{eqnarray}
In each single-bit-flip case, the resulting states can be perfectly decoded as long as the new encoding is learned.

We can learn which single bit flip has occurred without destroying the logical state by performing projective parity measurements $Z_{1}Z_{2}$ and $Z_{2}Z_{3}$ on the system, where the subscripts of the Pauli $Z$ operators indicate the bit number. These parity measurements give results $+1$ or $-1$ if the parity of the two coupled bits is even or odd, respectively. The parity measurements must be performed without measuring each qubit individually in order to preserve the coherence of the logical state. After performing a \emph{syndrome measurement} of the pair of parities $(Z_1 Z_2,\, Z_2 Z_3)$, we can use the \emph{syndrome} outcomes to identify the new logical encoding according to the mapping:
\begin{equation}\label{eq:syndromes}
\begin{array}{cc}
(+1,\,+1)\to & III,\\
(-1,\,+1)\to & XII,\\
(-1,\,-1)\to & IXI,\\
(+1,\,-1)\to & IIX.
\end{array}
\end{equation}
These syndrome measurements are checked periodically to detect single bit flips and infer the updated logical encoding. If desired, one could apply the operation of the encoding label to restore the encoding to the original encoding. For example, if we detect the parity measurement outcome is $(-1,\,+1)$, we know the encoding is $XII$; therefore, applying the operation $XII$ restores the encoding $III$ since applying $XII$ twice on the initial state yields the identity. However, this correction step may be delayed or omitted, since knowledge of the encoding is sufficient to use the coherent quantum information. Therefore, we assume passive error tracking, rather than active error correction, for the remainder of the paper.

\begin{figure}
\begin{centering}
\includegraphics[width=\columnwidth]{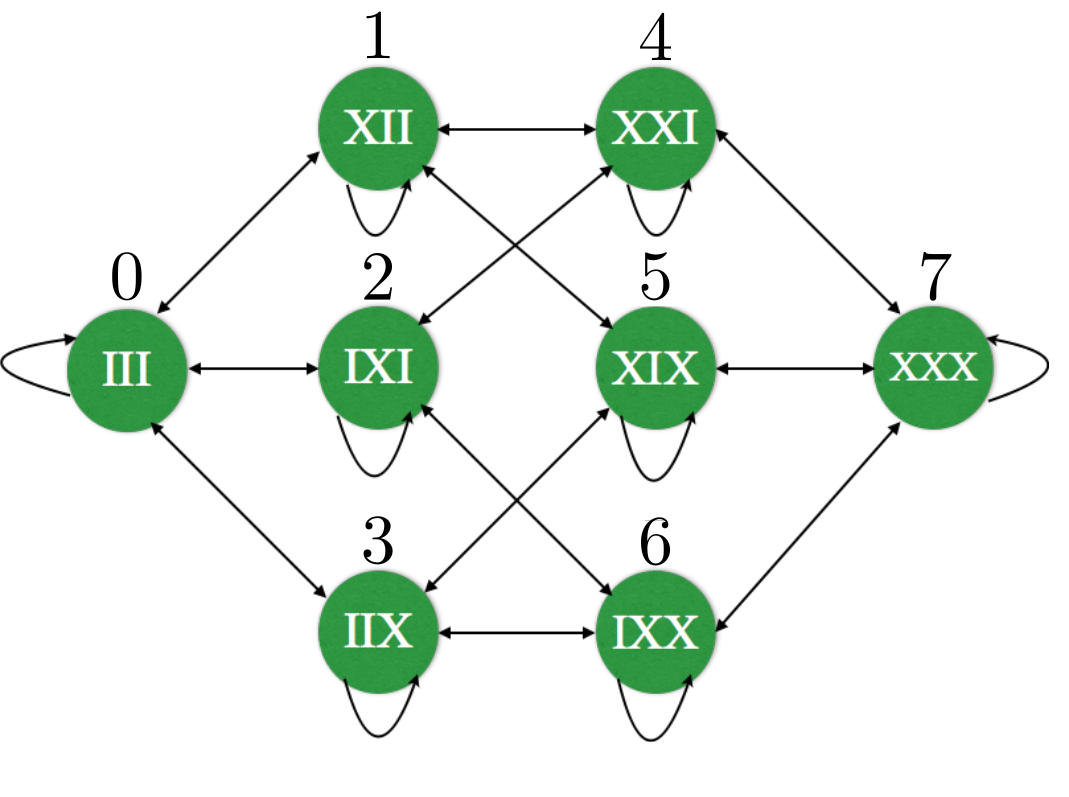}
\par\end{centering}
\caption{Hidden Markov model for the transitions between the eight logical encodings for the 3-qubit bit-flip code. Each encoding is labeled by the Pauli $X$ operations that relate it to the reference encoding $|\psi_0\ra = \alpha|000\ra + \beta|111\ra$, as well as a numeric index $k=0,\ldots,7$. Single bit-flips $X$ on each qubit cause transitions between encodings. Complementary encodings have identical parities, so cannot be distinguished by the syndrome measurements $(Z_1Z_2,\,Z_2Z_3)$, with bits numbered left-to-right as $123$. We assume that bit flips are independent and infrequent, with a constant rate $\mu$ per qubit.}
\label{fig:Markov}
\end{figure}

The code does not protect against two simultaneous bit flips from the $III$ encoding, denoted $XXI$, $XIX$, and $IXX$, which produce the states
\begin{eqnarray}\label{eq:encode5}
|\psi_4\ra &=& XXI|\psi_0\ra = \beta |001\ra + \alpha |110\ra, \\
|\psi_5\ra &=& XIX|\psi_0\ra = \beta |010\ra + \alpha |101\ra, \\
|\psi_6\ra &=& IXX|\psi_0\ra = \beta |100\ra + \alpha |011\ra.
\end{eqnarray}
Parity measurements of complementary bit states are identical, so the error syndromes will not correctly identify the change in encoding if two bit flips occur between two syndrome measurements. An incorrect identification of the encoding produces a \emph{logical error} since the quantum information can no longer be correctly decoded. The situation is the same with three bit flips, denoted $XXX$, which produces an encoding complementary to the original encoding
\be\label{eq:encode8}
|\psi_7\ra = XXX|\psi_0\ra = \beta|000\ra + \alpha |111\ra.
\ee
This syndrome ambiguity is not resolved by including a third parity measurement of $Z_{1}Z_{3}$, so we restrict our analysis to two parity measurements to minimize hardware resources. However, we note that adding the third parity measurement would slightly improve our ability to discriminate sequential bit flips from a single bit flip.

The code also does not protect against non-bit-flip errors of the data qubits, such as phase flips, which can also produce logical errors. Similarly, the code is not fault-tolerant, so does not protect against all errors that can appear during syndrome measurements, such as bit flips of ancillary qubits in the middle of an entangling gate. 

Our task is to track the transitions between the 8 encodings produced by bit flips, starting from the initial encoding $III$. We measure the syndromes to update our knowledge of the encoding. At some later time $t$, if we still know the correct encoding then we have tracked all bit flip errors successfully and thus can correctly decode the state. However, if we incorrectly track the encoding, then we have failed to track bit-flip errors, so trying to decode the state will produce a logical error. 

We define the (binary) \emph{fidelity} $f(t)\in\{0,1\}$ of error tracking after a duration $t$ to be 1 if the knowledge of the encoding matches the true encoding, and 0 if they differ. The \emph{average fidelity} $F(t)\in[0,1]$ is the average of the binary fidelity over many tracking realizations---equivalent to the process fidelity in quantum process tomography---and serves as a useful performance metric. 

\subsection{Bit-flip error model and fidelity}\label{sec:code:errormodel}
For simplicity of analysis, we assume that bit flips occur independently, infrequently, and at a slow but constant rate $\mu$ per qubit, so that the flips are Poisson-distributed in time. We take the bit-flip rate to be equal for each of the three qubits for simplicity and symmetric with respect to the bit states. To focus on the flipping dynamics, we work in the rotating frame of the physical qubits, which remain uncoupled, so the effective idling Hamiltonian is zero. 

With these assumptions, the bit-flip-tracking task reduces to finding the evolution of a hidden Markov model \cite{zucchini2017hidden,gales2008theapplication,vanhandel2005optimal,mabuchi2009continuous}, with the possible transitions illustrated as the arrows in Fig.~\ref{fig:Markov}. Each encoding $k=0,\ldots,7$ described in the previous subsection has a probability $P_k\in[0,1]$ such that $\sum_k P_k = 1$. We assume the initial encoding state is $III$. The master equation that describes the jump processes on average can then be expressed as a matrix equation
\begin{align}\label{eq:mastereqmatrix}
  \partial_t \vec{P} &= \mathbf{M}\, \vec{P},  & P_0(0) &= 1,
\end{align}
with probability vector $\vec{P} = [P_0\, P_1\, P_2\, P_3\, P_4\, P_5\, P_6\, P_7]^T$ and Markov transition matrix 
\begin{align}\label{eq:markovmatrix}
  \mathbf{M} &= \mu\begin{bmatrix} -3 & 1 & 1 & 1 & 0 & 0 & 0 & 0 \\ 1 & -3 & 0 & 0 & 1 & 1 & 0 & 0 \\ 1 & 0 & -3 & 0 & 1 & 0 & 1 & 0 \\ 1 & 0 & 0 & -3 & 0 & 1 & 1 & 0 \\ 0 & 1 & 1 & 0 & -3 & 0 & 0 & 1 \\ 0 & 1 & 0 & 1 & 0 & -3 & 0 & 1 \\ 0 & 0 & 1 & 1 & 0 & 0 & -3 & 1 \\ 0 & 0 & 0 & 0 & 1 & 1 & 1 & -3 \end{bmatrix}.
\end{align}
Note that we neglect double-flip or triple-flip processes in this matrix. The solution $\vec{P}(t) = \exp(t\mathbf{M})\,\vec{P}(0)$ asymptotically approaches the uniform distribution as a fixed point for large $t$, $\lim_{t\to\infty} P_k = 1/8$. 

The average encoding fidelity $F(t) \equiv P_0(t)$ may be obtained by solving Eq.~\eqref{eq:mastereqmatrix}. 
Each bit flips independently, so the solution factors into a product of exponential decays of each bit to an asymptotic flip probability of $1/2$. The average fidelity with no jump tracking is thus
\begin{align}\label{eq:fidnotrack}
	F(t) \equiv P_0(t) &= \left[\frac{1 + \exp(-2\mu t)}{2}\right]^3 \\
&= 1 - 3\mu t + 6 \mu^2 t^2 + \cdots. \nonumber
\end{align}
The fractional deviation of this decay from the linear regime is $(6\mu^2 t^2)/(3\mu t) = 2\mu t$. For later optimizations we bound this deviation by $1/15$ to ensure that the linear approximation $F(t) \approx 1 - 3\mu t$ is a reasonable decay model, which bounds $\mu t \leq 1/30$ and thus the maximum average fidelity drop while remaining in the linear regime to $1-F(t) \leq 10\%$. 

For practical error-tracking purposes, it is sufficient to focus on improving the short-time fidelity with linear decay by tracking the jumps with syndrome measurements. After including jump tracking, the approximate form of the fidelity in the linear regime will be
\begin{align}\label{eq:lineardecay}
  F(t) = 1 - \Delta F_{\text{in}} - \Gamma t,
\end{align}
where $\Delta F_{\text{in}}$ is the \emph{initial drop in fidelity} on a short time scale, while $\Gamma$ is the average \emph{logical error rate} for longer time scales after the tracking method takes full effect. 

In later sections we will derive approximate expressions for and optimize this linear drop in fidelity to assess the relative performance between error correction methods. We will find that with good error correction the optimized linear decay $\Gamma$ scales as $\mu^2$ after the short-duration initial drop in fidelity $\Delta F_{\text{in}}$ that is still linear in $\mu$. 


\section{Physical Setup}\label{sec:setup}

\begin{figure}[t]
\begin{centering}
\includegraphics[width=\columnwidth]{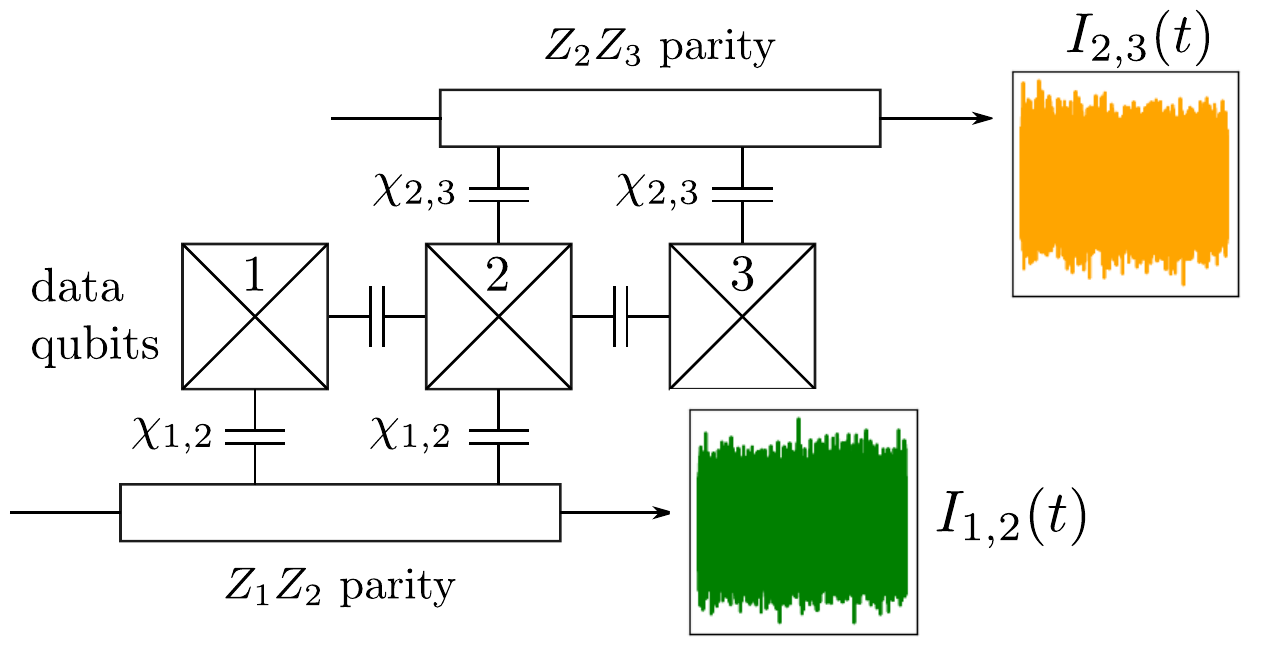}
\par\end{centering}
	\caption{Possible experimental setup for continuous bit flip error correction using the three-bit code. The parity of neighboring qubits $(z_i z_j) = \pm 1$ is measured directly by coupling both qubits to the same readout resonator such that they each dispersively shift the resonator frequency by the same amount, $\chi_{i,j}$. Pumping on resonance then populates the resonator field entangled with the odd parity subspace, leaving the even parity subspace entangled with a near-vacuum state. Homodyne measurement produces a stochastic signal, $I_{i,j}(t)$, that directly reveals the parity after renormalization, $I_{i,j}(t) \to r_{i,j}(t) = (z_i z_j)(t) + \sqrt{\tau}\,\xi_{i,j}(t)$, with $\la \xi_{i,j}(t)\xi_{i,j}(0) \ra = \delta(t)$. Integrating this stochastic signal for the characteristic time scale $\tau$ produces a unit signal-to-noise ratio for identifying the qubit-qubit parity. 
\label{fig:Experimental-setup}}
\end{figure}

The goal of a physical realization of the three-qubit code is to perform the syndrome measurements of the two-qubit parities $(Z_1Z_2,\,Z_2Z_3)$ and use them to track bit flips to preserve the knowledge of the logical state encoding. We focus on direct syndrome measurement that is continuous in time, in contrast to traditional ancilla-based periodic syndrome measurements. For concreteness, we consider one possible physical implementation with modern superconducting transmon qubits \cite{Koch2007} on a two-dimensional wafer, shown in Fig.~\ref{fig:Experimental-setup}, where the parities are measured continuously via dispersive coupling to microwave resonators \cite{LivingstonAPS2018,LivingstonAPS2019}. While a third parity measurement of $Z_1Z_3$ is possible in principle by wrapping the qubits into a ring, it increases the complexity of the hardware.

In this configuration, the readout resonators are coupled to pairs of data qubits to directly measure the parity. The dispersive shifts $\chi$ for each qubit (e.g., $\chi_{1,2}$ or $\chi_{2,3}$ in Fig.~\ref{fig:Experimental-setup}) must be tuned to be identical, such that they are comparable to or greater the linewidth $\kappa$ of the resonator, $2\chi \gg \kappa$. By fabricating qubits 1 and 3 with tunable SQUID loops, the dispersive shifts can be tuned to match as required \cite{LivingstonAPS2018,LivingstonAPS2019}. The odd-parity subspace with two-qubit states $|01\ra$ and $|10\ra$ will shift the resonator frequency first up by $\chi$ then down by $-\chi$ (or vice versa) to return to its original resonance frequency. The even-parity subspace will have $|11\ra$ shift the frequency by $2\chi \gg \kappa$ while $|00\ra$ will shift by $-2\chi$ so that the line widths do not overlap strongly. Hence, the resonant pump will produce a non-vacuum steady-state field in the resonator only for the odd-parity subspace, leaving the even-parity subspace in vacuum. 

The parity subspaces therefore become entangled with two distinct coherent fields: $(c_{00}|00\ra + c_{11}|11\ra + c_{01}|01\ra + c_{10}|10\ra)|\alpha=0\ra \to (c_{00}|00\ra + c_{11}|11\ra)|\alpha=0\ra + (c_{01}|01\ra + c_{10}|10\ra)|\alpha = \beta\ra$ with $|\beta|>0$. This entanglement enables homodyne measurement of the leaked resonator field to distinguish the subspaces. The coherence of each subspace, however, remains essentially unperturbed because the fields for each parity subspace are indistinguishable within the subspace. Realistically, imperfect field overlap can still dephase the parity subspaces, which is an imperfection analogous to entangling-gate infidelity in ancilla-based parity-measurements. For simplicity of analysis, we assume this dephasing is sufficiently slow to neglect. 

After amplifying the leaked fields and measuring them via homodyne detection along the maximally informative field quadrature, a stochastic signal is obtained for each parity resonator. The resonator connected to data qubits 1 and 2 produces the signal $I_{1,2}(t)$, while the resonator connected to data qubits 2 and 3 produces the signal $I_{2,3}(t)$. After properly shifting and normalizing these signals, they approximate moving-mean Gaussian stochastic processes centered at the parity eigenvalues $(z_iz_j) = \pm 1$:
\begin{equation}\label{eq:paritysignals}
\begin{array}{cc}
dr_{1,2}(t) &= (z_1z_2)(t)\,dt + \sqrt{\tau_{1,2}}\,dW_{1,2},\\ 
dr_{2,3}(t) &= (z_2z_3)(t)\,dt + \sqrt{\tau_{2,3}}\,dW_{2,3}.  
\end{array}
\end{equation}
Here, $dW_{1,2}$ and $dW_{2,3}$ are statistically independent Wiener increments, each with zero mean Gaussian statistics and variance $dt$. Formally these increments can also be understood as $\delta$-correlated white noise, $\xi_{i,j} \equiv dW_{i,j}/dt$, with $\langle \xi_{i,j}(t)\xi_{i,j}(t')\rangle = \delta(t-t')$. For simplicity in what follows, we assume both noises are characterized by the same \emph{characteristic measurement timescale} $\tau_{1,2} = \tau_{2,3} = \tau$, which signifies the integration duration needed to achieve unit signal-to-noise ratio (SNR). The parity information can thus be recovered by processing the stochastic signal over a duration of time. As a temporal reference in simulations, we will fix the measurement timescale to be fast, $\tau=100$ ns, and consider relatively slow bit-flip rates in the range $\mu\tau\in[10^{-6},\,10^{-3}]$. 

This direct parity-measurement method reduces hardware resources compared to an ancilla-qubit-based approach. Such a gate-based approach would require two additional ancilla qubits to measure the parities (in addition to the readout resonators for each ancilla qubit), as well as periodic entangling gates and projective measurements. In contrast, the direct parity-measurement method considered here requires only a single readout resonator per parity measurement. The direct method also yields a raw, time-continuous parity signal, which can be processed in two distinct ways for the purposes of error correction: 
\begin{enumerate}
   \item Continuous filtering to track the most likely errors that have occurred in real time
   \item Periodic filtering by integrating and thresholding over consecutive durations $\Delta t$
\end{enumerate}	
Notably, the second method can use the same error-tracking algorithm as for the ancilla-based approach with periodic projective measurements. We now analyze both methods in the following sections.

%
%
\section{Continuous Bayesian filter}\label{sec:continuous}

Environmental perturbations during monitoring cause jumps between the encoding states in Fig.~\ref{fig:Markov}. Encodings connected by a single jump have distinct parity eigenvalues, so the means of the noisy parity signals in Eqs.~\eqref{eq:paritysignals} will correspondingly jump. Integrating these noisy signal with a moving temporal window with a duration longer than $\tau$ can therefore identify infrequent single jumps \cite{Slichter2016}, allowing the changes in logical encoding to be tracked via the changing syndromes. However, if multiple jumps occur within a time scale comparable to $\tau$, then the noise can prevent the jumps from being identified before the encoding jumps to a complementary one with a parity indistinguishable from the original one. Such a misidentification of an encoding with its complement oding is a logical error that will not be corrected by continued monitoring. It is thus important to filter the noisy signals in a way that minimizes misidentification errors caused by rapid successive jumps. 

An optimal time-continuous filter can be derived by using all available information to process the time-continuous noisy signals. The key idea is to update the encoding probabilities $\vec{P}$ at each moment in time using Bayes' rule, which requires known likelihoods of observing the collected signals given definite parities and a known estimate of the flipping rate $\mu$. The maximum resulting probability then indicates the best guess for the updated encoding. 

Importantly, the fidelity of tracking the encoding is determined only by the correctness of the estimate at the final time. The assumed Markovian dynamics imply that each random jump and random noise fluctuation is independent of past fluctuations, which implies that adding information from temporally extended signal correlations will not improve the final state estimate. In particular, time-symmetric smoothing methods \cite{Simonoff1998,Ellison2009,Einicke2012}) that process the past signal still produce estimates identical to forward-in-time estimates for the state at the final time (as we have verified numerically), even though they do generally improve the tracking fidelity for past jumps. It is thus sufficient to consider only forward-in-time Bayesian updates to derive a filter that uses all relevant information about the stochastic signal and the flipping dynamics to achieve an optimal state estimate at the final time. 

Such a time-continuous Bayesian filter is known as a Wonham filter \cite{wonham1964someapplications}, and has been applied to continuous error correction of the three-bit code by van Handel and Mabuchi \cite{vanhandel2005optimal,mabuchi2009continuous}. However, the Wonham filter contains a nonlinear update from Bayes' rule that reduces its computational efficiency during real time processing of the stochastic signals. To address this problem, we introduce a variation of the Wonham filter that removes this nonlinearity to improve computational efficiency. Our \emph{linear Wonham filter} uses unnormalized probabilities $\vec{\sigma}(t)$ that reproduce the correct probabilities after renormalization $\vec{P}(t) = \vec{\sigma}(t)/||\vec{\sigma}(t)||_1$ with the 1-norm $||\vec{\sigma}(t)||_1 = \sum_{k=0}^7 \sigma_k(t)$. These unnormalized probabilities can be regularized periodically only as needed, drastically reducing the computational overhead of real-time processing with the filter. We expect this linear filter to be suitable for real-time processing with FPGAs to enable on-demand state estimation and feedback. We derive and analyze this filter in what follows.

\subsection{Derivation of linear Bayesian filter}\label{sec:continuous:bayes}

Recall that the Markovian master equation for the encoding probabilities $\vec{P}(t)$ without error tracking is Eq.~\eqref{eq:mastereqmatrix}. The goal is to update this evolution to include the information gained by the stochastic parity measurements. This new information will refine the probability evolution with Bayes' rule.

Before deriving the linear filter, we first derive the nonlinear Wonham filter for comparison. The deterministic dynamics of the bit-flips is unchanged by the probabilistic updates from the measurement results, so the contribution of the averaged master Eq.~\eqref{eq:mastereqmatrix} will be unchanged in the final dynamical equation. For this reason, we will initially neglect this deterministic part in the derivation, then add it back at the end.

\subsubsection{Nonlinear Bayesian (Wonham) filter}
After averaging the stochastic signals over a short duration $dt$, the rescaled readouts $\bar{r}_{1,2}$ and $\bar{r}_{2,3}$ for the two continuous parity measurements are Gaussian with independent noises according to Eqs.~\eqref{eq:paritysignals}, so the joint probability density of both results is a product of Gaussian distributions,
\begin{align}
  P(\bar{r}_{1,2}, \bar{r}_{2,3}\, |\, k) &= P(\bar{r}_{1,2}\,|\,k)P(\bar{r}_{2,3}\,|\,k), \\
  P(\bar{r}_{i,j}\,|\,k) &= \frac{\exp(-dt(\bar{r}_{i,j} - s_{i,j|k})^2/2\tau)}{\sqrt{2\pi\tau/dt}}.
\end{align}
Here the index $k = 0, \ldots, 7$ indicates a definite encoding as described in Fig.~\ref{fig:Markov}, and the means $s_{i,j|k} = \pm 1$ are the parity eigenvalues of the encoding shown in Table~\ref{tab:parities}.

\begin{table}[t]
 \begin{center}
 \begin{tabular}{ c|r@{\hspace{1em}}|r@{\hspace{1em}} } 
  \;$k$\; & \; $s_{1,2|k}$  &\;  $s_{2,3|k}$  \\ 
  \hline \hline
  0 & +1 & +1 \\ 
  1 & -1 & +1 \\ 
  2 & -1 & -1 \\ 
  3 & +1 & -1 \\
  4 & +1 & -1 \\ 
  5 & -1 & -1 \\ 
  6 & -1 & +1 \\ 
  7 & +1 & +1 \\ 
 \end{tabular}
 \end{center}
\caption{Parity eigenvalues $s_{i,j|k}$ for each encoding $k$.}
\label{tab:parities}
\end{table}

After collecting integrated readouts, each encoding probability $P_k$ in $\vec{P}$ should be updated via Bayes' rule,
 \be\label{eq:bayesrule}
 P_k \xrightarrow{(\bar{r}_{1,2},\,\bar{r}_{2,3})} \frac{P(\bar{r}_{1,2},\, \bar{r}_{2,3}\,|\, k) P_k}{\sum_\ell P(\bar{r}_{1,2},\, \bar{r}_{2,3}\,|\,\ell)P_\ell}.
 \ee
Since the likelihood probabilities are Gaussian with means that always square to 1, this update ratio considerably simplifies to
 \be\label{eq:bayesgauss}
 P_k \to \frac{\exp\left[(dt/\tau)(\bar{r}_{1,2}s_{1,2|k} + \bar{r}_{2,3}s_{2,3|k})\right] P_k}{\sum_\ell \exp\left[(dt/\tau)(\bar{r}_{1,2}s_{1,2|\ell} + \bar{r}_{2,3}s_{2,3|\ell})\right]P_\ell}.
 \ee

This update is already sufficient to track the most likely state given a temporal sequence of integrated readouts $\{\bar{r}_{i,j}(n\, dt)\}_{n=0}^N$. However, it can be conceptually useful to put the time-continuous deterministic evolution of Eq.~\eqref{eq:mastereqmatrix} on equal footing with the Bayesian updates by taking the time-continuous limit of the latter to produce a \emph{filtering equation} that includes both stochastic and deterministic updates. To do this, we expand the Bayesian update equation to first order in $dt$ to obtain a nonlinear stochastic differential equation (SDE) in Stratonovich form (with time-symmetric derivative obeying standard calculus rules), to which we can simply add the deterministic part of the evolution from Eq.~\eqref{eq:mastereqmatrix} giving 
\begin{align}\label{eq:wonham}
	& \text{(Stratonovich)} \\
\partial_t P_k &= \sum_\ell M_{k\ell}P_{\ell} + (\delta_{k\ell} - P_k)\,P_\ell\, \frac{s_{1,2|\ell}\,r_{1,2} + s_{2,3|\ell}\,r_{2,3}}{\tau}, \nonumber
\end{align}
where $M_{k\ell}$ are the components of the transition matrix $\mathbf{M}$ in Eq.~\eqref{eq:markovmatrix}, and $\delta_{k\ell}$ is the Kronecker delta.
This equation can be used directly to convert the data stream into a state estimation.

\begin{figure*}[t]
\begin{center}
\includegraphics[width=1.4\columnwidth]{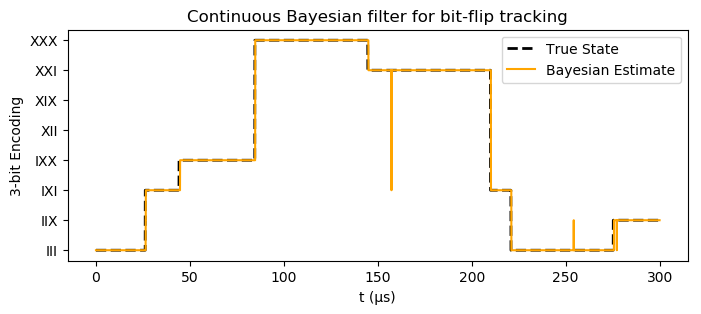} \\
\includegraphics[width=1.4\columnwidth]{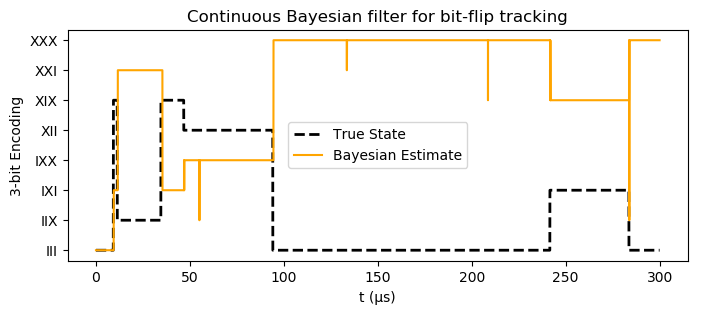}
\end{center}
\caption{Two examples of the linear Bayesian filter used for bit-flip tracking. The initial encoding is $III$ (see Fig.~\ref{fig:Markov}), with characteristic measurement time of $\tau = 0.1\mu$s, and a bit-flip rate of $\mu = 5\times 10^{-3}\,(\mu\text{s})^{-1}$. (top) Successful tracking. Single bit flips are identified after a brief delay comparable to $\tau$. Three filter errors caused by noise fluctuations are shown at times $160\mu$s, $255\mu$s, and $275\mu$s, which are all quickly self-corrected. (bottom) Unsuccessful tracking due to a logical error. Two bit flips (bits 1 and 3) occur in rapid succession at $10\mu$s, faster than the time scale of $\tau$ can detect. The filter incorrectly interprets this pair as a bit 2 flip, which is a logical encoding error. The filter never recovers, and continues tracking the complementary encoding.}
\label{fig:Bayesian}
\end{figure*}
The stochastic process can be modeled by converting the SDE to It\^o form with a forward-difference derivative, which modifies the equation by adding an effective drift term \cite{Gardiner1997}.
After lengthy calculation, the added drift cancels the means of the stochastic signals $r_{i,j}(t) = (z_iz_j)(t) + \sqrt{\tau}\,\xi_{i,j}(t)$ to leave only the zero-mean white noise $\xi_{i,j}(t) \equiv dW_{i,j}(t)/dt$,
\begin{align}\label{eq:wonhamito}
	& \text{(It\^o)} \\
\partial_t P_k &= \sum_\ell M_{k\ell}P_{\ell} + (\delta_{k\ell} - P_k)\,P_\ell\, \frac{s_{1,2|\ell}\,\xi_{1,2} + s_{2,3|\ell}\,\xi_{2,3}}{\tau}. \nonumber
\end{align}
Since in It\^o form the noise terms $\xi_{i,j}$ are uncorrelated with each other and with the state probabilities at the earlier time step, this form of the equation makes it clear that averaging over all noise realizations eliminates the stochastic terms, leaving just the drift, to correctly recover the original Lindblad form master Eq.~\eqref{eq:mastereqmatrix} without tracking. This SDE is the nonlinear Wonham filter used by van Handel and Mabuchi \cite{wonham1964someapplications,mabuchi2009continuous}.

\subsubsection{Linear Bayesian filter}
We will now linearize the Wonham filter by removing the nonlinear normalization step from the Bayesian update. To do this, we define ``unnormalized probabilities'' $\vec{\sigma}$ such that $\vec{\sigma}/||\vec{\sigma}||_1 = \vec{P}$ recovers the same encoding probabilities as before. We then modify the key Bayesian update step of Eq.~\eqref{eq:bayesgauss} by omitting the denominator:
 \be\label{eq:bayesgausslinear}
 \sigma_k \xrightarrow{(\bar{r}_{1,2},\,\bar{r}_{2,3})} \exp\left[\frac{dt}{\tau}(\bar{r}_{1,2}s_{1,2|k} + \bar{r}_{2,3}s_{2,3|k})\right] \sigma_k.
 \ee
Note that we have preserved the cancellation of state-independent Gaussian factors in the Bayesian update to prevent irrelevant (state-independent) changes in the norm. This \emph{linearized} update isolates only the state-dependent changes to the unnormalized probabilities.

Proceeding as before, we can expand this update to linear order in $dt$ to obtain a filtering equation. Importantly, the instantaneous signals $r_{i,j}(t) = (z_i z_j)(t) + \sqrt{\tau}\,\xi_{i,j}(t)$ depend only on the definite parity $(z_i z_j)(t) = \pm 1$ of the actual encoding (i.e., not an expectation value in an estimated state), so do not depend upon the distinction between $\vec{P}$ or $\vec{\sigma}$. We then add the deterministic updates as before with one modification: we remove the diagonal part $-3\mu\mathbf{I}$ of the transition matrix in Eq.~\eqref{eq:markovmatrix} that is proportional to the identity matrix $\mathbf{I}$. Any such term proportional to the identity causes irrelevant increases in the norm of $\vec{\sigma}$ and thus can be added or removed arbitrarily without affecting the relative sizes of its components. This freedom of choice is analogous to choosing a gauge and will be useful in the derivation to follow. The Stratonovich filtering equation then takes the simple linear form
\begin{align}\label{eq:linearbayes}
	\partial_t \vec{\sigma} &= \left(\widetilde{\mathbf{M}} + \frac{r_{1,2}\mathbf{S}_{1,2} +r_{2,3}\mathbf{S}_{2,3}}{\tau}\right)\,\vec{\sigma},
\end{align}
where
\begin{align}
	\label{eq:markovmatrixmod}
	\widetilde{\mathbf{M}}_{k\ell} &\equiv (1-\delta_{k\ell})\,\mathbf{M}_{k\ell}, \\
	(\mathbf{S}_{i,j})_{k\ell} &\equiv \delta_{k\ell}\,s_{i,j|k}.
\end{align}
This \emph{linear Bayesian filter} directly depends on the measured signals $r_{i,j}(t)$ as well as the diagonal matrices $\mathbf{S}_{i,j}$ of the parities $s_{i,j|k}$ shown in Table~\ref{tab:parities}. Converting this equation to It\^o form simply adds a state-independent drift term, $\vec{\sigma}/\tau$, on the right hand side that only changes the overall norm and can thus be omitted. As a result, Eq.~\eqref{eq:linearbayes} can be understood in either the Stratonovich or It\^o picture without changing the results that will be predicted by the maximum unnormalized probability,
\be
k_{\text{est}} \equiv \text{argmax}_k\,\sigma_k.
\ee

We show examples of this filter being used for error tracking in Fig.~\ref{fig:Bayesian}. We contrast one example trial with only single bit-flip errors (top) against one trial with one ``double-flip'' error that is uncorrectable (bottom). In the successful tracking case, the filter tightly tracks the actual encoding jumps, with a delay in detecting jumps set by the characteristic measurement time $\tau$. Noise fluctuations cause occasional errors that are rapidly corrected on the same time scale of $\tau$. In the unsuccessful tracking case, two successive jumps that occur on a timescale faster than $\tau$ are misinterpreted as a different single bit flip, which produces a logical error that the filter is not designed to correct.

\subsection{Bayesian filter analysis}\label{sec:continuous:analysis}
We now derive simple expressions for the linear degradation of state-tracking fidelity in Eq.~\eqref{eq:lineardecay}. We first consider the initial fidelity drop $\Delta F_{\text{in}}$, then consider the linear decay rate $\Gamma$ at steady-state after the filter takes full effect. The Bayesian filter has no free parameters to optimize; it only depends on knowledge of the parity eigenvalues, the characteristic measurement time $\tau$ of the collected signals, and the estimated bit-flip rate $\mu$. As a result, the derived expressions for fidelity decay provide an estimate for the best tracking fidelity that could be achieved in principle with continuous parity measurements. 

\subsubsection{Initial fidelity drop}
The Bayesian filter has an initial drop in fidelity $F_{\text{in}}$ primarily because of its delayed response to a bit flip. This delay makes the filter vulnerable to bit flips that occur just before the final state estimate is requested. We thus expect a drop in fidelity by the probability of a bit flip occurring within one filter response time.  We stress this is a general feature of all error correction techniques and is in no way special to our protocol.

The filters starts from time $t=0$ with the correct encoding. Since the initial encoding is 0 $(III)$ with certainty, we focus on the encodings reachable by one bit flip: 1 ($XII$), 2 ($IXI$), and 3 ($IIX$) according to the numbering in Fig.~\ref{fig:Markov}. For simplicity, we initially neglect the Gaussian noise in Eq.~\eqref{eq:linearbayes} to focus on the evolution caused by the signal means. We also use the freedom of the norm to add a constant term to Eq.~\eqref{eq:linearbayes} and shift the parity eigenvalues after each jump to 0 for the correct states and -2 for incorrect states; this shift simplifies the analysis by keeping correct state (unnormalized) probabilities nearly constant. 

Focusing on the four states relevant from the previous paragraph $(\sigma_0, \sigma_1, \sigma_2, \sigma_3)$, the equation of motion is
\begin{equation}
    \partial_t {\vec \sigma} = \begin{pmatrix}
    0 & \mu & \mu & \mu \\
    \mu & -2/\tau & 0 & 0\\
     \mu & 0 & -4/\tau & 0\\
      \mu & 0 &  0 & -2/\tau 
      \end{pmatrix} \cdot {\vec \sigma}.
\end{equation}
Here we have assumed that the true state of the system is $(III)$, so the parities read by the detectors is $\langle r_{1,2} \rangle= 1, \langle r_{2,3} \rangle = 1$ (even,even).  From Eq.~(\ref{eq:linearbayes}) the diagonal term is thus $\{1, -1, -1, 1\}/\tau + \{1, 1, -1, -1\}/\tau = \{2, 0, -2, 0\}/\tau$.  We have used the freedom of the overall norm of $\vec \sigma$ to subtract a factor of $2{\bf I}/\tau$ from all diagonal entries, so the correct state (0) does not grow, but rather the incorrect states decay.

It is now straightforward to see that starting from the initial condition $(1, 0, 0, 0)$ the components $\sigma_1$ and $\sigma_3$ reach a steady state of $\mu \tau/2$, while the component $\sigma_2$ reaches its steady state of $\mu \tau/4$.  The true component $\sigma_0$ actually grows very slowly from 1 as $\sigma_0 \approx 1 + (5/4)\mu^2 \tau t$, but we can neglect this correction on the time scales of interest.  This is then the idling state of the filter while in the error-free original state.

Suppose now that a bit flip occurs, either $XII$, $IXI$, or $IIX$.  We take first a flip on bit 1.  The parity eigenvalues then change immediately to the values $\langle r_{1,2} \rangle= -1, \langle r_{2,3} \rangle = 1$ (odd,even).  Consequently, the filter equation changes the diagonal term to $-\{1, -1, -1, 1\}/\tau + \{1, 1, -1, -1\}/\tau = \{0, 2, 0, -2\}/\tau$, and we again shift the overall matrix by $-2{\bf I}/\tau$ to get the new equation
\begin{equation}
    \partial_t {\vec \sigma} = \begin{pmatrix}
    -2/\tau & \mu & \mu & \mu \\
    \mu & 0 & 0 & 0\\
     \mu & 0 & -2/\tau & 0\\
      \mu & 0 &  0 & -4/\tau 
      \end{pmatrix} \cdot {\vec \sigma},
\end{equation}
where we now have the initial condition just found, $(1, \mu \tau/2, \mu \tau/4, \mu \tau/2)$.  These equations are readily solved to find $\sigma_0 = e^{-2 t/\tau}$ and $\sigma_1(t) = \mu \tau - (\mu \tau/2) e^{-2 t/\tau}$, which quickly limits to its new steady state of $\mu \tau$.

The filter is able to catch the error when the value of $\sigma_1$ exceeds the value of $\sigma_0$.  This timescale defines the {\it response time} of the filter.  Solving then $\mu \tau =  e^{-2 t/\tau}$, we find the response time of the filter to a qubit 1 flip to be $t_r^{(1)} = (\tau/2) \ln (1/\mu \tau)$.  The definition of the {\it initial drop} of the filter fidelity is if the bit flip occurs before the filter can respond appropriately.  Consequently, if the process is called after an error occurs, but before the filter can respond, then a logical error happens.  The probability of this occurring is the drop in fidelity, $\Delta F = \mu t_r^{(1)}$, which is linear in $\mu$.

Repeating this analysis for qubit 3 gives the same result, $t_r^{(3)}= (\tau/2) \ln (1/\mu \tau)$.  For a qubit 2 error, both parities change (odd,odd), so the relevant equation of motion is
\begin{equation}
    \partial_t {\vec \sigma} = \begin{pmatrix}
    -4/\tau & \mu & \mu & \mu \\
    \mu & -2/\tau & 0 & 0\\
     \mu & 0 & 0 & 0\\
      \mu & 0 &  0 & -2/\tau 
      \end{pmatrix} \cdot {\vec \sigma},
\end{equation}
because the diagonal term is now $-\{1, -1, -1, 1\}/\tau - \{1, 1, -1, -1\}/\tau = \{-2, 0, 2, 0\}/\tau$, with the overall subtraction.
Starting from the same initial conditions as before, the solutions are $\sigma_0(t) = e^{-4 t/\tau}$, and $\sigma_2(t) = (\mu \tau/2) - (\mu \tau/4)e^{-4 t/\tau}$, which limits quickly to $\mu \tau/2$.  Thus, the time when $\sigma_0$ becomes smaller than $\sigma_2$ is given by $t_r^{(2)} = (\tau/4) \ln(2/\mu \tau)$.
 
Adding up the drop contributions for the three single bit flips produces
\be\label{eq:bayesiandrop}
\Delta F_{\text{in}} = \mu\tau\left[\frac{5}{4}\ln\left(\frac{1}{\mu\tau}\right) + \frac{1}{4}\ln 2\right].
\ee
The initial drop in fidelity is linear in $\mu$ up to logarithmic corrections, since the error correction has not taken full effect. As a reminder, we expect the long-time decay after error correction takes effect to be quadratic in $\mu$.

This estimate for the initial drop has neglected the role of the Gaussian noise in the signals. As seen in Fig.~\ref{fig:Bayesian}, noise fluctuations can occasionally cause the filter to jump to a different state estimate even when no bit flip occurs. These fluctuations produce false positives that are usually quickly corrected on the time scale of the filter response and do not contribute to the logical error rate. However, if such a fluctuation occurs just before the termination time, then the false positive is not corrected before the estimated state is requested, resulting in a misidentification error. A drop contribution from such noise-induced misidentifications should be added to the flip-based drop estimate in Eq.~\eqref{eq:bayesiandrop}.

As we present in Section~\ref{sec:simulation}, we have numerically checked Eq.~\eqref{eq:bayesiandrop} with flip rates ranging from $\mu\tau \in [10^{-6},10^{-3}]$. We found that the derived expression systematically underestimates the drop by a small amount, as anticipated from the omission of the noise contribution. In order to correct this systematic underestimation in a crude way, we found that it is sufficient to alter the numerical prefactor in the first term of Eq.~\eqref{eq:bayesiandrop} by substituting $5/4 \mapsto 3/2$, effectively adding a noise-based contribution that is approximately half that expected from a flip-induced single parity flip $\mu t^{(1)}_r/2 = (\mu\tau/4)\ln(1/\mu\tau)$. Though physically unjustifiable due to the lack of true bit flip, this adjustment compensates for the additional noise-based drop and agrees with numerics within the range $\mu\tau \in [10^{-6},10^{-3}]$. A proper treatment of the noise-induced drop is analytically lengthy and beyond the scope of the simple derivations given here, so we make this crude prefactor substitution for simplicity in the plots of Section~\ref{sec:simulation}.

\subsubsection{Logical error rate of the Bayesian filter}

In addition to the errors contributing to the initial drop in fidelity, which only occur just before the final time, the Bayesian filter is vulnerable at any time to logical errors caused by two consecutive bit flips that occur within one response time of the filter. Since the first bit flip does not have time to be registered by the filter, the two flips will be interpreted as a single flip, which causes the filter to track an incorrect complementary encoding, as shown in the bottom half of Fig.~\ref{fig:Bayesian}. These logical errors require two flips, so produce a logical error rate that scales as $\mu^2$.

Logical errors of this type can be produced by 6 double-flip scenarios, which we can reduce to three distinct cases by symmetry. The sequence of bit 1 flipping then bit 2 flipping, (i.e., $\{1,2\}$) produces the same error as the $\{3,2\}$ flip sequence. Similarly, the $\{2, 1\}$ and $\{2, 3\}$ sequences produce identical errors, as do the $\{1 ,3\}$ and $\{3, 1\}$ sequences. We thus consider only three distinct cases: $\{1,2\}$, $\{2,1\}$, and $\{1,3\}$.

We start with the $\{1, 2\}$ case. Consider a bit 1 flip at time $t=0$ (chosen arbitrarily) followed by a bit 2 flip at a later time $T$ that is faster than the filter can resolve it.   After the second bit flip the correct encoding is $4$ ($XXI$).  The filter can sometimes make an error and return the value of $3$ ($IIX$), because it cannot resolve the time between the two parity flips, and can only see the transition from $0$ to $3$, connected by a single bit flip on the third qubit (see Fig.~\ref{fig:Markov}). Formally, this occurs if $\sigma_4(t) < \sigma_3(t)$ asymptotically for $t\gg T$.  We will now calculate the rate at which this mistake can occur, which leads to a logical error that can not be corrected.

We generalize the analysis of the last subsection by also including the dynamics of state $\sigma_4$, which will be the true state at the end of the section.  This state connects to states 1 and 2 by bit flips, and before the first flip has the equation of motion $\partial_t \sigma_4 = \mu(\sigma_1+\sigma_2) - 2\sigma_4/\tau$.  Its steady state value is $3\mu^2\tau^2/8$.

Once qubit 1 flips at time $t=0$, the new diagonal terms of the equation matrix are associated with (odd,even) parities, and become $\{-2,0,-2,-4,-4\}/\tau$.  The equation of motion for state 4 is now $\partial_t \sigma_4 = \mu(\sigma_1+\sigma_2) - 4\sigma_4/\tau$.  The steady-state value is $\sigma_4 = \mu^2\tau^2/4$.  The results obtained before for $\sigma_0, \sigma_1$ still hold, and we find the solution $\sigma_3(t)= (\mu \tau/2) e^{-2t/\tau}$.  For later convenience, we also note that solutions for the other components are $\sigma_2(t) = \mu t e^{-2 t/\tau}$ and $\sigma_5(t) = \mu^2\tau^2/2$ in the steady state.

Qubit 2 then flips at time $t=T$, making the true state of the system now 4 ($XXI$) and the parity eigenvalues (even, odd). The new diagonal terms of the equation matrix become $\{-2,-4,-2,0,0\}/\tau$.  This indicates that neither state 3 nor state 4 are dynamically suppressed by the filter, because both have the correct parities associated with them.  We now reset our time, and need to solve the following approximate set of equations
\begin{align}
\partial_t \sigma_0 &= -2\sigma_0/\tau,\\
\partial_t \sigma_1 &= -4\sigma_1/\tau, \nonumber\\
\partial_t \sigma_3 &= \mu \sigma_0, \nonumber\\
\partial_t \sigma_4 &= \mu \sigma_1, \nonumber
\end{align}
starting with the initial conditions $\sigma_0(T) = e^{-2T/\tau}, \sigma_1(T) = \mu \tau, \sigma_3(T) = (\mu \tau/2) e^{-2T/\tau}, \sigma_4(T) = \mu^2 \tau^2$. In the equations above, we have kept the leading order terms assuming $\mu \tau \ll 1$ (states 2,5,7 are not relevant to this discussion).  These equations can be solved with standard methods, leading to the asymptotic results for $t \gg T$ of ${\bar \sigma}_3 = \mu \tau e^{-2 T/\tau}$, and ${\bar \sigma}_4 = \mu^2 \tau^2/2$.

The filter will give a logical error if ${\bar \sigma}_3 > {\bar \sigma}_4$ since it returns the incorrect state.  From the asymptotic results derived above, this error occurs when
\begin{equation}
    T < \frac{\tau}{2} \ln\left(\frac{2}{\mu \tau}\right).
\end{equation}
This result makes physical sense:  If the second error occurs at a time shorter than the filter response time, it cannot sense the difference between the two scenerios we have sketched here.  The {\it logical error rate} is the rate at which this kind of process occurs.  This rate can be calculated as the rate of the first error occurring, $\mu$, times the probability that a second error occurs within a time $T$ after it, $\mu T$.  Consequently, the error rate is given by $\Gamma_{\{1,2\}} = (\mu^2\tau/2) \ln(2/\mu \tau)$. 

We next consider the $\{2, 1\}$ case. The reverse scenario of a bit 2 flip followed by a bit 1 flip produces a very similar derivation to the one above, which we omit for brevity, and also yields the same condition $T < (\tau/2)\ln(2/\mu\tau)$. Therefore, the contributions to the logical error rate from this scenario (or bit 2 then bit 3 flips) are the same $\Gamma_{\{2,1\}} = \Gamma_{\{2,3\}} = (\mu^2\tau/2)\ln(2/\mu\tau)$.

Finally, we consider the $\{1, 3\}$ case. The scenario of a bit 1 flip followed by a bit 3 flip produces a slightly different result than the other two cases. After the bit 1 flip, at time $T$ we have the same states from before, the largest of which are $\sigma_0(T) = \exp(-2T/\tau)$ and $\sigma_1(T) = \mu\tau$. Now, qubit 3 flips instead of qubit 2, so the relevant states are 
 5 ($XIX$), the correct state, and 2 ($IXI$), the incorrect one, with the same parity results of (odd,odd).   The relevant diagonal terms in the matrix equation are now $\{-4, -2, 0, -2, -2, 0\}/\tau$, so we now focus on the $\sigma_0, \sigma_1, \sigma_2, \sigma_5$ dynamics.  The equations of motion for this situation are then,
 \begin{align}
\partial_t \sigma_0 &= -4\sigma_0/\tau,\\
\partial_t \sigma_1 &= -2\sigma_1/\tau, \nonumber \\
\partial_t \sigma_2 &= \mu \sigma_0, \nonumber \\
\partial_t \sigma_5 &= \mu(\sigma_1 +\sigma_3), \nonumber
\end{align}
with the initial conditions established after the qubit 1 bit flip, $\sigma_0(T) = e^{-2 T/\tau}$, $\sigma_1(T) = \mu \tau$, $\sigma_2(T) = \mu T e^{-2T/\tau}$, and $\sigma_5(T) = \mu^2 \tau^2/2$.  Solving these equations with standard methods yields for $t \gg T$ the asymptotic results, ${\bar \sigma}_5 = \mu^2 \tau^2$, and   ${\bar \sigma}_2 = \mu T e^{-2 T/\tau}$.  We can find the logical error rate by finding the time where ${\bar \sigma}_2 >{\bar \sigma}_5$, or whenever $t <T$, where
\begin{equation}
    T = \frac{\tau}{2} \ln \frac{T}{\mu \tau^2} \approx \frac{\tau}{2} \ln\left[ \frac{2}{\mu \tau} \ln \frac{c}{\mu \tau} \right],
\end{equation}
 and $c$ is a number of order 5 in the logarithm approximation.
 
By symmetry, all other error processes identified in the beginning of the section reduce to the type identified above.  Adding them yields the total logical error rate,
\begin{align}\label{eq:bayesiangamma}
  \Gamma = \mu^2\tau \left[3\,\ln\frac{2}{\mu\tau} + \ln\frac{\ln(5/\mu\tau)}{4}\right].
\end{align}
We numerically verify this expression over the range of bit-flip rates $\mu\tau\in[10^{-6},10^{-3}]$ in Section~\ref{sec:simulation}.

\section{Periodic Filters}\label{sec:periodic}

The linear Bayesian filter analyzed in the preceding section produces an optimal estimate and is more computationally efficient than the nonlinear Wonham filter. However, it requires prior knowledge of the bit-flip rate $\mu$ and the Gaussian noise time scale $\tau$, and still requires several matrix multiplications per time step. We wish to compare this optimal case against simpler and more practical filters that require less prior information and are more easily implementable in hardware, e.g. with field-programmable gate arrays (FPGAs), to enable on-demand state estimation for purposes of feedback control. We consider variations of a particularly simple ``boxcar-averaging'' filter, which should be well-suited for low-latency hardware. 

Boxcar filters average successive durations $\Delta t$ of the noisy parity signals $r_{i,j}(t)$, then threshold the integrated means $\bar{r}_{i,j} = \int_0^{\Delta t}r_{i,j}(t)dt/\Delta t$, often using two thresholds $a_+$ and $a_-$, with $a_+>a_-$. That is, if the average signal exceeds the threshold $a_+$, we assign the value $+1$ to the parity. Similarly, if the time-averaged signal is less than the threshold $a_-$, then we assign the value $-1$ to the parity. Any result between the two thresholds $a_+$ and $a_-$ is treated as ambiguous, mandating a separate strategy for resolving the ambiguity. The final outputs of the filter are binary parity results, $(b_{1,2},\,b_{2,3})$ with $b_{i,j} = \pm 1$, for each time duration $\Delta t$. These results can then be used to track changes in the encoding using the syndromes in Eq.~\eqref{eq:syndromes}. This filter is the most direct translation of standard ancilla-based error correction with periodic projective measurements to continuous syndrome measurements. The primary difference is that continuous measurements are always on, and the collected data is only later partitioned into bins of duration $\Delta t$ for averaging and tracking.

In what follows, we analyze three simple boxcar-averaging filter variations:
\begin{enumerate}[A.]
	\item Boxcar filter : The simplest method for averaging sequential time bins of duration $\Delta t$, using symmetric thresholds $a_+=a_-=0$.
	\item Half-boxcar filter : A non-Markovian modification to the simple boxcar filter that removes its dominant source of error by occasionally processing the averaged signal shifted by a half duration $\Delta t/2$.
	\item Double-threshold boxcar filter : A Markovian modification to the simple boxcar filter that also removes its dominant source of error, using asymmetric thresholds $a\equiv a_+\geq 0$ and $a_-=0$.
\end{enumerate}
We find that although the simplest boxcar filter performs poorly compared to the Bayesian filter, the two proposed modifications can achieve performance comparable to the optimal Bayesian filter with significantly less computational overhead. Note that for these considered filters, only two tunable parameters must be set prior to filtering: the boxcar duration $\Delta t$ for all three filters, and the asymmetric threshold $a\geq 0$ for the double threshold filter.  

Over the next few subsections, we identify the dominant error mechanisms and define the three boxcar filters in more detail. For each filter, we derive expressions for the long-time linear decay rate $\Gamma$. We then derive expressions for their initial fidelity drops $\Delta F_{\text{in}}$, since they arise from similar mechanisms. We then consider optimization of the tunable filter parameters and analytically optimize the parameters to obtain simpler formulas that can be directly compared with those of the Bayesian filter. We numerically verify both the optimal parameters and the derived expressions in Section~\ref{sec:simulation}.

\subsection{Boxcar error mechanisms}
There are two main mechanisms for causing a change in syndrome in a boxcar filter: (i) A bit flip can occur with rate $\mu$ (yielding a probability of flip per averaging box of $\mu\Delta t$), which can alter one or both parities. (ii) Noise fluctuations can cause the average parity signal over a box to appear changed, even though no actual bit flip occurs. Such a misidentification of a parity flip will be incorrectly interpreted by the filter as an actual bit flip. 

Logical errors are produced by sequences of these basic mechanisms occurring at particular times. For example, a bit flip that occurs in the latter half of an averaging box will not produce a detectable parity flip until the subsequent box, which allows time for a second bit flip to occur and place the bits in a state complementary to the estimated state tracked by the filter. Similarly, if one parity is misidentified in an averaging box, a nearby bit flip can confuse the filter so that it tracks an estimation that is complementary to the actual state. We detail these dangerous event sequences in the following sections.

Parity misidentification errors play an important role in the following analysis, so we give a general analysis of their probabilities to occur here. A parity misidentification occurs when the integrated signal for a box is observed to be less than the discrimination threshold $a$, even though no bit flip occurs. Given an integration duration $\Delta t$ and mean parity $r_m$ over that duration (e.g., $r_m = \pm 1$ for a definite parity that persists the entire duration), the probability of obtaining an integrated signal $\bar{r}$ is Gaussian, $P(\bar{r}\,|\,r_m) = \exp(-(\bar{r}-r_m)^2\Delta t/2\tau)/\sqrt{2\pi\tau/\Delta t}$. Thus, the probability of obtaining an integrated signal less than a discrimination threshold $a$ is
\begin{align} \label{eq:Idisc}
  P(\bar{r} < a\, |\, r_m) &= \int_{-\infty}^a \!\! P(\bar{r}\, |\, r_m)d\bar{r} \nonumber \\ 
  &= \text{erfc} \left[(r_m-a)\sqrt{\Delta t/2 \tau}\right]/2, 
\end{align}
where $\text{erfc}(x) = 1-\text{erf}(x)$ is the complementary error function. 

The probability of misidentifying a parity of $+1$ as $-1$ is therefore
\begin{align}\label{eq:pmis2}
  P_{\text{mis}}(a) &\equiv P(\bar{r}<a\,|\,{+1}),
\end{align}
while misidentifying $-1$ as $+1$ has probability 
\begin{align}
  P(\bar{r}>a\,|\,{-1}) = P(\bar{r}<-a\,|\,{+1}) = P_{\text{mis}}(-a),
\end{align}
using the simplification $P(-\bar{r}\,|\,{-1}) = P(\bar{r}\,|\,{+1})$. For the simple boxcar case, when the threshold is the symmetric point $a=0$, these formulas simplify to a single misidentification probability,
\begin{align} \label{eq:pmis}
P_{\text{mis}} &\equiv P_{\text{mis}}(0), \\
& = \text{erfc}(\sqrt{\Delta t / 2\tau})/2, \nonumber \\
& \approx \exp(-\Delta t/2\tau)\sqrt{\tau/2 \pi \Delta t}. \nonumber
\end{align}
The final asymptotic exponential approximation is valid when $\Delta t \gg \tau$. Note that for reasonably long integration times $\Delta t \sim 10\tau$, the misidentification probability is less than 0.1\%.

\subsection{Boxcar logical error rate}\label{sec:periodic:box}

The simplest boxcar filter with symmetric threshold $a=0$ is an important reference case for the other boxcar filter variations. As such, we first analyze its error mechanisms in detail so that we can identify its dominant error. The subsequent boxcar variations will use different strategies to target and correct this dominant error. We focus here on deriving the dominant contributions to the logical error rate $\Gamma$, and delay the consideration of the initial drop $\Delta F_{\text{in}}$ until after all three boxcar variations have been carefully defined. 

Since the 3-bit code is designed to protect against only a single bit-flip error, the most straightforward contributions to the logical error rate $\Gamma$ are from pairs of errors. For example, the two-flip sequence $III \to XII \to XIX$ can be misinterpreted as a single flip $III \to IXI$ to the complementary encoding, causing a logical error. These problematic error pairs can be broadly categorized into three groups: (a) two bit flips, (b) one bit flip and one parity misidentification, or (c) two parity misidentifications. The contributions that involve three or more basic errors are comparatively small, so we will neglect them in this analysis.

\begin{figure}[t]
\begin{center}
\includegraphics[width=\columnwidth]{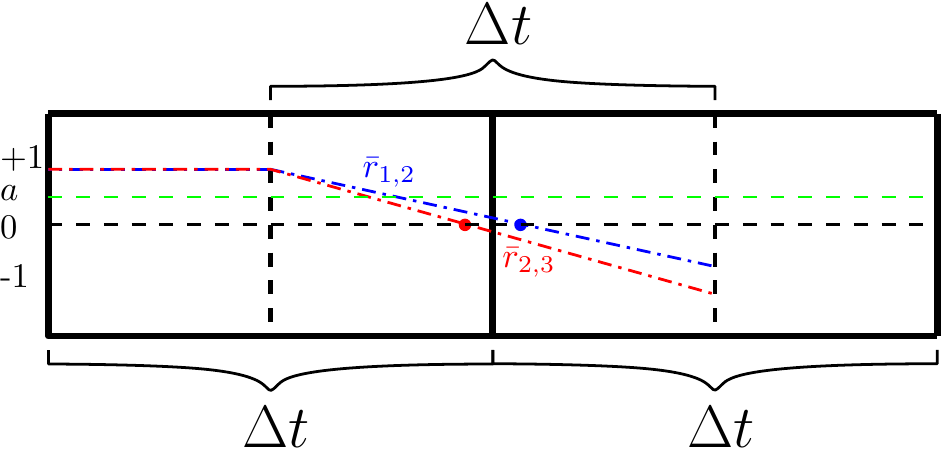}
\caption{
	Most significant boxcar-averaging error. If bit 2 of the three-bit code flips, then both normalized and averaged parity readouts $\bar{r}_{i,j}$ for bits i and j will flip over an averaging time $\Delta t$, as shown by the blue and red diagonal dot-dashed lines. When the flip occurs in the middle of an averaging box as shown, then the averaged readouts may cross the zero threshold at slightly different times due to noise fluctuations, causing one parity to appear flipped in one averaging box while the other appears flipped in the next averaging box. The sequence is interpreted as a succession of flips for bits 1 and 3, producing a logical error. We propose two variations to fix this error: (1) The non-Markovian half-boxcar method reevaluates the midsection between successive averaging boxes when bits 1 and 3 successively flip. If both parities flip for the midsection average, then the succession of flips is correctly reinterpreted as a bit 2 flip. (2) The Markovian double-threshold method introduces a second threshold $a\geq 0$, shown by the horizontal green dashed line, such that a flip in bit 2 is detected when both parity signals drop below this new threshold.
}\label{fig:boxcar}
\end{center}
\end{figure}

In addition to these mechanisms involving pairs of errors, however, there is a more subtle and dangerous single error mechanism: (d) a mid-box flip of bit 2. In this case a single flip $III \to IXI$ can cause the parities to flip in successive boxes due to their independent noise fluctuations. Negative-biased noise fluctuations can make one averaged parity pass the zero threshold within the first box, while positive-biased noise fluctuations can make the other parity pass the zero threshold at a later time that occurs in the subsequent box. As such, the reported parity flips will be interpreted as the sequence $III \to XII \to XIX$ and yield a complementary encoding, causing a logical error. Since this last type of error is caused by a single flip, it is the dominant source of error for the simple boxcar filter that will be removed by the boxcar variations in subsequent sections. We illustrate this problematic error mechanism in Fig.~\ref{fig:boxcar}, which we will refer to again when discussing the mechanisms of the boxcar variations that fix this error. 

We now discuss each error contribution in turn.
\begin{enumerate}[(a)]
  \item Two bit flips

  Each bit flip has an independent probability of $\mu\Delta t$. There are three ways to have two distinct flips with 3 qubits. Therefore, the probability for two distinct flips to occur within one box $\Delta t$ is $3(\mu \Delta t)^2$, yielding a contribution to the logical error rate of $\Gamma_\text{bb} = 3(\mu \Delta t)^2/\Delta t = 3\mu^2\Delta t$.

More precisely, a bit flip in the first half of a box is likely to be detected, but a bit flip in the second half is unlikely to be detected until the following box. The sensitivity region for flips is thus shifted by a half-box in time from the periodic syndrome information. That is, for two bit-flip errors the danger is in having two flips within a region of duration $\Delta t$ that starts at a mid point of one box and ends at the midpoint of the next box. However, this temporal shift by a half-box does not affect the reasoning used for the logical error rate.  

  \item One bit flip and one misidentification 

Logical errors generally require two averaging boxes to manifest. Over two consecutive boxes $2\Delta t$ there are two possible parities to misidentify in each box and three half-boxes $\Delta t/2$ in which a flip of one of the three bits could cause parity changes, yielding 36 total pairs of errors to consider. After checking each of these possibilities, we identify 4 classes that produce a logical error:

  \begin{enumerate}[i.]
    \item A misidentification and a flip of a complementary bit within the first half of the same box: For example, starting from encoding $III$ a misidentification in channel $r_{1,2}$ and a flip of bit 3 in the first half of the box produces a syndrome with two flipped parities, which is misinterpreted as a bit 2 flip. This error leaves the true encoding in $IIX$ while the estimated encodings follow the sequence $III \to IXI \to XXI$ to produce a logical error. There are 2 parities to misidentify in one box $\Delta t$, with 2 complementary bits each, so there are 4 possibilities of error, each with probability $\mu(\Delta t/2)\,P_{\text{mis}}$, producing a contribution to the logical error rate of $4(\mu\Delta t/2)\,P_{\text{mis}}/\Delta t = 2\mu\,P_{\text{mis}}$.
    
    \item A misidentification and a flip of a complementary bit within the second half of the same box: For example, misidentifying $r_{1,2}$ produces the estimated encoding $XII$. After a flip in bit 3 the true encoding becomes $IIX$, which in the following box will make it appear that both parities have flipped, leading to the estimated encoding sequence $III \to XII \to XXI$. As with the preceding case, there are 4 possibilities yielding a contribution to the error rate of $2\mu\,P_{\text{mis}}$.
    
    \item A misidentification in one box, then a bit flip in the first half of the next box: The parities from this sequence appear identical to the previous case. There are 4 possibilities, so the error rate contribution is also $2\mu\,P_{\text{mis}}$.
    
    \item A bit flip in the second half of a box, then a complementary parity misidentification in the following box: In the first box no error will be reported, but the second box will have two apparent bit flips interpreted as a wrong single flip. There are 4 possibilities, so the contribution is also $2\mu\,P_{\text{mis}}$.
  \end{enumerate}
  
  The total error rate contribution is $\Gamma_\text{bm} = 8 \mu P_{\text{mis}}$.

  \item Two misidentifications
 
  A misidentification in one channel followed by a second misidentification in the complementary channel during the next box causes a logical error. For example, misidentifications of $r_{1,2}$ then $r_{2,3}$ causes the estimated encoding sequence over three boxes: $III\to XII\to XXI \to XXX$. There are two orderings for this type of error, so the contribution to the logical error rate is $\Gamma_\text{mm} = 2 P_{\text{mis}}^2/\Delta t$.

  \item Mid-box flip of bit 2

  Starting with the encoding $III$, suppose that bit 2 flips near the center of a box, at time $\Delta t/2+\delta t$. Both parities should flip in this case, but the flip occurs in a region where the integration result will be sensitive to noise fluctuations. If one parity shows a flip while the other does not, it leads to a logical error over the course of two boxes. For example, if $r_{1,2}$ flips in one box while $r_{2,3}$ flips in the next box, the apparent encoding follows the sequence $III \to XII \to XIX$. This error is unique to a bit 2 flip because such a flip requires both parities to correctly flip.

The integrated parity signal $\bar{r}$ after such a flip is Gaussian-distributed with a mean value of $r_m = [(\Delta t/2 + \delta t) - (\Delta t/2 - \delta t)]/\Delta t = 2 \delta t/\Delta t$ and variance $\tau/\Delta t$. The probability of getting $\bar{r}<0$ in a channel is equal to $P(\bar{r}<0\,|\,r_m)$ as defined in Eq.~\eqref{eq:Idisc}. If one channel flips $\bar{r}_{1,2} < 0$, but the other does not $\bar{r}_{2,3}>0$, then the probability of this occurring for any $r_m\in[-1,1]$ is 
\begin{align}\label{eq:halfboxerror}
\int_{-1}^1 P(\bar{r}<0\,|\,r_m)\,P(\bar{r}>0\,|\,r_m)\,dr_m \to \sqrt{\frac{\tau}{\pi\Delta t}},
\end{align}
where the integral over the product of error functions rapidly reaches an exact asymptotic value after $\Delta t \gtrsim 10 \tau$. The contribution of this scenario to the logical error rate is therefore $\Gamma_\text{b2} = \mu\sqrt{\tau/\pi\Delta t}$. As highlighted previously, this is the most dangerous error mechanism because its error rate is linearly dependent on $\mu$.
\end{enumerate}

Gathering all of the above contributions produces the total logical error rate for the boxcar filter, arranged in order of significance: 
\begin{align}\label{eq:gammaboxcar}
  \Gamma &= \Gamma_\text{b2} + \Gamma_\text{bb} + \Gamma_\text{bm} + \Gamma_\text{mm}, \\
	&= \mu\sqrt{\frac{\tau}{\pi\Delta t}} + 3 \mu^2\Delta t +  8 \mu \, P_{\text{mis}} + 2\frac{P^2_{\text{mis}}}{\Delta t} \nonumber
\end{align}
We will later optimize this formula over the boxcar duration $\Delta t$ and numerically verify its accuracy over the range of bit-flip rates $\mu\tau \in [10^{-6},10^{-3}]$.

Because of the dominant error that is linearly dependent on $\mu$ the boxcar filter performs poorly compared to the Bayesian filter and is not useful for error correction. To make the boxcar filter viable, we therefore wish to eliminate this dominant error with a simple and minimal modification to the boxcar filter. We consider two such modifications in the subsequent sections: a non-Markovian modification that uses previous history in the continuous record to identify the problematic bit-2 flip, and a Markovian modification that identifies the problematic bit-2 flip by bracketing the parity signals between two thresholds. We will see that both variations compare favorably to the Bayesian method, so are suitable for practical error correction. 
 
\subsection{Half-boxcar filter and error rate}\label{sec:periodic:halfbox}

To overcome the problem of the bit-2 flip in the boxcar filter, we introduce a non-Markovian extension that we call the half-boxcar filter. To the basic boxcar filter we add one extra conditional action that reexamines any ostensible sequential flips of bits 1 and 3 to make sure they are not an incorrectly interpreted flip of bit 2. That is, if a parity flip is observed after an averaging boxcar, $t\in[0,\,\Delta t]$, and a second parity flip is observed in the opposite channel one boxcar later, $t\in[\Delta t,\,2\Delta t]$, then the filter reexamines the signal in an interval that straddles both boxcars. The raw signal is reaveraged over a duration $\Delta t$ that is shifted one half-boxcar behind the most recent boxcar, $t\in[\Delta t/2,\,3\Delta t/2]$, and compared to the zero threshold as a secondary check. 

With this modification, a flip in bit 2 that happens near the center of the first box will cause both parities to change in the re-averaged middle box, unlike sequential flips of bits 1 and 3. The top portion of Fig.~\ref{fig:boxcar} illustrates how both averaged signals will flip when averaging the shifted middle box in the protocol. Therefore, this modification correctly distinguishes a bit-2 flip from sequential bit-1 and bit-3 flips and eliminates the primary logical error mechanism of the boxcar filter. When a bit-2 flip is detected, the interpreted history of bit-flips must then be corrected so that the first box, $t\in[0,\,\Delta t]$ records correctly that bit 2 flipped, while the second box, $t\in[\Delta t,\,2\Delta t]$, records that nothing additional occurred.

An elegant implementation of this non-Markovian extension averages sequential half-box intervals $\Delta t/2$ of the raw signals, storing the most recent three half-box averages in memory in addition to the accumulating sequence of parity values. Averaging pairs of these pre-integrated half-boxes then efficiently produces either the most recent full-box average or the required shifted box to reassess a suspected bit-2 flip as needed. As such, this extension only minimally increases the computational complexity compared to the basic boxcar filter, while improving the fidelity so that it compares favorably with the Bayesian filter. It thus achieves a good balance between accuracy and efficiency. 
 
\subsubsection{Half-boxcar logical error rate}
We follow the same procedure for categorizing contributions to the logical error rate $\Gamma$ as we did for the boxcar filter in the preceding section. While the addition of the half-box mechanism removes the most serious of the boxcar errors, it also subtly alters the other logical error mechanisms, both removing a few more errors and adding new ones. We now discuss each category of contributions in turn.
\begin{enumerate}[(a)]
  \item Two bit flips

Unlike the basic boxcar filter, the partitioning of two consecutive boxes into half-boxes matters for sequences of two bit flips. Logical errors can occur from bit flips within the same box, or two consecutive boxes. There are four relevant cases.

    \begin{enumerate}[i.]
      \item Two distinct bit flips in the same half-box: There are three possibilities, each with probability $(\mu \Delta t/2)^2$. The contribution to the logical error rate is thus: $3 \, (\mu \Delta t/2)^2 \, (2/\Delta t) = (3/2)\, \mu^2 \Delta t$.
     	   
      \item  Two consecutive flips of bits 1 and 3, one in the second half of a box and the other in the first half of the following box: The first flip is not detected in the first box, so both parities will flip in the second box and be incorrectly interpreted as a flip in bit 2 after the second box. There are two possible orderings, so the total contribution to the logical error rate is: $2 \, (\mu\Delta t/2)^2/\Delta t = \mu^2 \Delta t/2$.
    
      \item A flip in either bit 1 or 3 during the second half of a box followed by a flip in bit 2 during the first half of the following box: The first flip is not detected after the first box, so only one parity will flip in the second box and be incorrectly interpreted as a flip of the complementary bit. After a third box, both parities will appear change and be incorrectly interpreted as a bit-2 flip, which leaves the estimate in a complementary state. There are two possibilities, with a similar situation if the bits flip in reverse order, so the total contribution to the logical error rate is  $4 \, (\mu\Delta t/2)^2 /\Delta t = \mu^2 \Delta t$.
      
      \item Two consecutive flips of bits 1 and 3, one in each half of the same box: The second flip is not detected, so one parity appears to flip, followed by the other parity in the next box. The half-box prescription then averages the middle of the boxes, which will show that both parities flip, and thus be misinterpreted as a flip in bit 2, which is a logical error. There are two possible orderings, so the total contribution to the logical error rate is $2 \,(\mu \Delta t/2)^2 /\Delta t = \mu^2 \Delta t/2$.
    \end{enumerate}
     
    The final error above is newly introduced by the half-boxcar mechanism, so the total contribution of two bit flips to the logical error rate is larger than the simple boxcar filter: $\Gamma_\text{bb} = (7/2) \mu^2 \Delta t$.
    
\item One bit flip and one misidentification

There are 6 distinct mechanisms for a bit flip and misidentification to cause a logical error. The half-boxcar filter modifies these contributions significantly from the simple boxcar filter.

	\begin{enumerate}[i.]
	\item A misidentification and a flip of a complementary bit both within the same box: This mechanism is the same as the boxcar case, but with the improvement that any flip in bit 2 is now corrected by the half-box mechanism. There are 2 possible misidentifications, each with 1 complementary bit that flips with probability $\mu\Delta t$. The total contribution to the logical error rate is $2\mu P_{\text{mis}}$.
    
 	\item A misidentification then a complementary flip of bit 1 or 3 during the first half of the next box: This appears as a sequence of two flips not corrected by the half-boxcar mechanism. For example, if bit 1 flips then the actual encoding becomes $XII$, but the apparent encoding follows the sequence $III \to IIX \to IXX$. There are two possibilities, so the contribution is $\mu P_{\text{mis}}$.
    
 	\item A misidentification then a complementary flip of bit 1 or 3 during the second half of the next box, but near the middle: The bit flip will be reported with probability $P(\bar{r}<0\,|\,r_m)$, with $r_m = 2\delta t/\Delta t$ as in Eq.~\eqref{eq:halfboxerror}, where $\delta t$ is the location of the flip with respect to the middle of the box, resulting in the same logical error as in the previous case. There are two possibilities, so the contribution to the logical error rate is $2\,\mu\,P_{\text{mis}}\int_0^1 P(\bar{r}<0\,|\,r_m)\,dr_m \to \mu\,P_{\text{mis}}\,\sqrt{\tau/2\pi\Delta t}$. This exact asymptotic value is reached by $\Delta t \gtrsim 15\tau$.
    
 	\item A misidentification then a flip in bit 2 during the first half of the next box: This produces an apparent sequence of flips in bits 1 and 3, which triggers the half-box mechanism. However, for the half-box-shifted middle section that is checked, bit 2 flipped too late to be detected. It is thus possible for only one parity to flip, analogously to the original bit-2 flip issue of the basic boxcar in Eq.~\eqref{eq:halfboxerror}, which leaves the logical error uncorrected. The contribution to the logical error rate is $2 \mu P_{\text{mis}}\int_0^1 P(\bar{r}< 0\,|\,r_m)P(\bar{r}>0\,|\,r_m)\,dr_m \to \mu\,P_{\text{mis}}\,\sqrt{\tau/4\pi\Delta t}$. This exact asymptotic value is reached by $\Delta t \gtrsim 15\tau$.
 		
 	 \item A misidentification then a flip in bit 2 during the second half of the next box: This scenario appears identical to the preceding case, so also contributes $\mu\,P_{\text{mis}}\,\sqrt{\tau/4\pi\Delta t}$.
 	
 	\item A flip in bit 2 near the middle of a box that triggers the half-box mechanism, followed by a misidentification in one of the channels during the check of the middle box: The check will then not correct the misinterpretation of the bit 2 flip as two consecutive bit 1 and bit 3 flips. The probability of this occurring is identical to the preceding two cases, so also contributes $\mu\,P_{\text{mis}}\,\sqrt{\tau/4\pi\Delta t}$.
	\end{enumerate}
    
	The total contribution to the logical error rate is $\Gamma_\text{bm} = 3\mu P_{\text{mis}} + (1 + 3/\sqrt{2})\, \mu P_{\text{mis}}\, \sqrt{\tau/2\pi\Delta t}$. 

\item Two misidentifications

The mechanism for two misidentification to cause a logical error is unchanged from the boxcar filter, so contributes $\Gamma_\text{mm} = 2 P_{\text{mis}}^2/\Delta t$.
\end{enumerate}
 
Gathering all contributions produces the total logical error rate for the half-boxcar filter:
\begin{align}\label{eq:gammahalfbox}
	\Gamma &= \Gamma_\text{bb} + \Gamma_\text{bm} + \Gamma_\text{mm} \\
	&= \frac{7}{2}\, \mu^2\Delta t + 3\mu\, P_{\text{mis}} + \frac{\sqrt{2}+3}{2}\sqrt{\frac{\tau}{\pi\Delta t}}\, \mu P_{\text{mis}}  + 2\frac{P^2_{\text{mis}}}{\Delta t}. \nonumber
\end{align}
Note that the value of the prefactors for several terms is achieved only for $\Delta t \gtrsim 15\tau$. We will later optimize the free parameter $\Delta t$ to find that this condition is satisfied self-consistently for $\mu\tau \lesssim 10^{-4}$, and verify this expression numerically for the bit-flip rates $\mu\tau \in [10^{-6},10^{-4}]$, with slight numerical deviations visible for $\mu\tau \in [10^{-4},10^{-3}]$ due to shorter optimal $\Delta t$ violating the approximation of the prefactor integrals.

\subsection{Double-threshold filter and error rate}\label{sec:periodic:thresh}
The non-Markovian half-boxcar filter has the drawback of requiring extra memory and reinterpreting the past tracking record. We thus also introduce an alternative filter that also corrects the problem of the bit 2 flip in the simple boxcar filter while remaining Markovian. This new filter saves on memory at the expense of extra conditional processing per boxcar.

We use the intuition that if a bit-2 flip happens near the center of a box, then both integrated parities should be near zero, with only noise fluctuations determining their sign about the usual threshold of zero. However, if a succession of bit 1 and bit 3 flips happens, then only one parity will cross the threshold at a time. We thus use a second signal threshold $a > 0$, that together with the zero threshold can bracket a region that checks whether both parities are simultaneously close to zero. Fig.~\ref{fig:boxcar} demonstrates this effect with the horizontal dashed green line for $a > 0$: Both averaged signals enter the region between $0$ and $a$ during the first boxcar, making the bit-2 flip correctly detectable using the second threshold.

More precisely, assuming an initially even-parity encoding $III$, if both integrated signals are less than the new threshold $a$, then we infer that bit 2 has likely flipped. Otherwise, the signals are thresholded as normal. In pseudocode, given an estimated encoding $III$,
\texttt{
\begin{verse}
if $\bar{r}_{1,2}<a$ and $\bar{r}_{2,3}<a$ \\
\qquad then flip bit 2\\
elseif $\bar{r}_{1,2}<0$ \\
\qquad then flip bit 1\\
elseif $\bar{r}_{2,3}<0$ \\
\qquad then flip bit 3\\
else \\
\qquad do nothing
\end{verse}
}
\noindent
where the flips are performed on the estimated state in accordance with passive error tracking. More generally, for an initial estimated encoding with parities $P_{1,2},\,P_{2,3} \in \{+1,-1 \}$, the parity-corrected integrated signals $(\bar{r}_{1,2}P_{1,2})$ and $(\bar{r}_{2,3}P_{2,3})$ should be used in the above algorithm in place of $\bar{r}_{1,2}$ and $\bar{r}_{2,3}$. The relaxed threshold can more robustly detect simultaneous parity changes when bit 2 flips close to the middle of a box, while remaining Markovian. 

\subsubsection{Double-threshold logical error rate}
One more we follow the same procedure as the boxcar filter to find the remaining contributions to the logical error rate $\Gamma$. As with the half-boxcar filter, the additional correction mechanism alters the mechanisms for producing logical errors. We now consider each category of contributions in turn.
\begin{enumerate}[(a)]
\item Two bit flips

The logical error rate caused by two bit flips in the same box is exactly the same as the basic boxcar filter: $\Gamma_\text{bb} = 3 \mu^2 \Delta t$. 
 
\item One misidentification and one bit flip

There are three distinct contributions:
\begin{enumerate}[i.]
	\item A zero-threshold misidentification, then a complementary flip in the next box: This case is the same as the basic boxcar, so the contribution of this kind of error to the logical error rate is $4\, \mu\, P_{\text{mis}}$.
    
 	\item An $a$-threshold misidentification, then a complementary flip of bit 1 or 3 in the same box: Since both parities are observed to be below $a$, this is interpreted by the double-threshold filter incorrectly as a bit 2 flip. There are two possibilities, so the contribution to the logical error rate is $2\, \mu\, P_{\text{mis}}(a)$, recalling the general form of the misidentification probability from Eq.~\eqref{eq:pmis2}.
    
 	\item A flip in bit 2 near the middle of a box, at time $\Delta t/2 + \delta t$, followed by parity misidentification: As with the bit-2 flip in the boxcar case, the parity signals will both be Gaussian-distributed with mean $r_m = 2 \delta t/\Delta t$ and variance $\tau/\Delta t$ as in Eq.~\eqref{eq:halfboxerror}. The probability of misidentifying the parity requires $\bar{r}_{1,2} > a$ while $\bar{r}_{2,3} < 0$, or vice versa. There are two possibilities, so the contribution to the logical error rate after summing over all $r_m\in[-1,1]$ is 
\begin{align}\label{eq:doublethresherror}
&2\mu\,\int_{-1}^1 P(\bar{r}<0\,|\,r_m)\,P(\bar{r}>a\,|\,r_m)\,dr_m \\
&\; \approx 2\mu\,\sqrt{\frac{\tau}{\pi\Delta t}}\,\exp\left[-0.9\,a\,\sqrt{\frac{\Delta t}{\tau}} - 0.15\,a^2\,\frac{\Delta t}{\tau}\right]. \nonumber
\end{align}
This Gaussian approximation to the error function integral is very accurate for $\Delta t \gtrsim 10\tau$ and $0\leq a \leq 1$, and correctly reduces to Eq.~\eqref{eq:halfboxerror} when $a=0$. It can be derived using the approximation $\text{erfc}(x) \approx \exp(-c_1x - c_2x^2)$ valid for $x>0$ with $c_1 \approx 1.1$ and $c_2 \approx 0.76$ \cite{Tsay2014}.
 	\end{enumerate}
    
The total contribution to the logical error rate is therefore $\Gamma_\text{bm} = 4 \mu P_{\text{mis}}+ 2 \mu P_{\text{mis}}(a) + 2\mu\,\sqrt{\tau/\pi\Delta t}\,\exp\left[-0.9\,a\,\sqrt{\Delta t/\tau} - 0.15\,a^2\,\Delta t/\tau\right]$.

\item Two misidentifications

The second threshold slightly modifies the simple boxcar contribution. After one misidentification of probability $P_{\text{mis}}$, either bit 1 or bit 3 flips. The next box corrects this error unless a second misidentification in the complementary channel occurs to make it appear that both parities have flipped. However, it is sufficient for both integrated signals to be less than $a$ in this case, due to the double threshold mechanism. Thus with probability $P_{\text{mis}}(a)$ there is a flip in bit 2. In the third box the remaining bit will flip, producing a logical error. There are two possibilities, so the contribution to the logical error rate is $\Gamma_\text{mm} = 2\, P_{\text{mis}}\, P_{\text{mis}}(a)/\Delta t$.
\end{enumerate}
  
Gathering all above contributions produces the total logical error rate:
\begin{align}\label{eq:gammadouble}
  \Gamma &= \Gamma_\text{bb} + \Gamma_\text{bm} + \Gamma_\text{mm}, \\
	&= 3\mu^2\tau \frac{\Delta t}{\tau} + 4 \mu\, P_{\text{mis}} + 2\, \mu\, P_{\text{mis}}(a) + 2\frac{P_\text{mis}\,P_{\text{mis}}(a)}{\Delta t} \nonumber \\
  &\quad  {} + 2\,\mu\,\sqrt{\frac{\tau}{\pi\Delta t}}\,\exp\left[-0.9\,a\,\sqrt{\frac{\Delta t}{\tau}} - 0.15\, a^2\, \frac{\Delta t}{\tau}\right]. \nonumber
\end{align}
We will later optimize this formula over the free parameters, the boxcar duration $\Delta t$ and threshold $a\geq 0$, and numerically verify its accuracy over the range of bit-flip rates $\mu\tau \in [10^{-6},10^{-3}]$.

\begin{table*}[t]
\small
\setlength{\tabcolsep}{8pt}
\renewcommand{\arraystretch}{4}
\begin{tabular}{l|l|l}
\multicolumn{1}{c|}{\textbf{Filter}} & $\Delta F_{\textrm{in}}$ & $\Gamma \tau$ \\
\hline\hline
Bayesian & $\displaystyle \mu\tau\left[\frac{5}{4}\ln\frac{1}{\mu\tau} + \frac{1}{4}\ln2 \right]$ & $\displaystyle 3(\mu\tau)^2\left[\ln\frac{2}{\mu\tau} + \frac{1}{3}\ln\frac{\ln(5/\mu\tau)}{4}\right]$ \\
\hline
Boxcar & $\displaystyle \frac{3\mu\tau}{2}\frac{\Delta t}{\tau}$ & $\displaystyle \mu\tau\, \sqrt{\frac{\tau}{\pi\Delta t}} + 3\, (\mu\tau)^2\, \frac{\Delta t}{\tau} +  8\, \mu\tau \, P_{\text{mis}} + 2\,P^2_{\text{mis}}\, \frac{\tau}{\Delta t} $  \\
\hline
Half-boxcar & $\displaystyle \frac{3\mu\tau}{2}\frac{\Delta t}{\tau} - \frac{\mu\tau}{2}\sqrt{\frac{\Delta t}{\pi\tau}} + \frac{\sqrt{2}\,e^{-\Delta t/2\tau}}{\sqrt{\pi \Delta t/\tau}}$ & $\displaystyle \frac{7}{2}\, (\mu\tau)^2\, \frac{\Delta t}{\tau} + 3\, \mu\tau\, P_{\text{mis}} + \left[\frac{1}{\sqrt{2}} + \frac{3}{2}\right]\sqrt{\frac{\tau}{\pi\Delta t}} \, \mu\tau\, P_{\text{mis}}  + 2\,P^2_{\text{mis}}\, \frac{\tau}{\Delta t}$\\
\hline
\begin{minipage}[t]{0.2\columnwidth}Double-\\threshold\end{minipage} & $\displaystyle \frac{3\mu\tau}{2}\frac{\Delta t}{\tau}$ & $\displaystyle \begin{aligned} & 3\, (\mu\tau)^2\, \frac{\Delta t}{\tau} + 4\, \mu\tau\, P_{\text{mis}} + 2\, \mu\tau\, P_{\text{mis}}(a) + 2\,P_\text{mis}\,P_{\text{mis}}(a)\,\frac{\tau}{\Delta t}
  \\ &\qquad  {} + 2\,\mu\tau\,\sqrt{\frac{\tau}{\pi\Delta t}}\,\exp\left[-0.9\,a\,\sqrt{\frac{\Delta t}{\tau}} - 0.15\, a^2\, \frac{\Delta t}{\tau}\right]\end{aligned}$ 
\end{tabular}
\caption{Initial drops in fidelity $\Delta F_{\text{in}}$ and logical error rates $\Gamma$ for various filters. We express the formulas in terms of the dimensionless quantities $\mu\tau$ and $\Delta t/\tau$, where $\mu$ is the bit-flip rate, $\tau$ is the measurement timescale, and $\Delta t$ is the averaging timescale for boxcar filters. The parity misidentification probabilities for the boxcar filters are $P_{\text{mis}}(a) \equiv \text{erfc}[(1-a)\sqrt{\Delta t/2\tau}]/2$ and $P_{\text{mis}} \equiv P_{\text{mis}}(0)$. We find numerically that for the Bayesian initial drop, more accurate results can be obtained in the regime $\mu\tau \in [10^{-6},10^{-3}]$ via the substitution $5/4 \mapsto 3/2$, which crudely compensates for neglected noise-fluctuation contributions. The three boxcar filters should be optimized over the averaging duration $\Delta t$, while the double-threshold variation should also be optimized over the threshold $a\geq0$. For the initial drop of the boxcar filters, we have approximated that $\Delta t/\tau \gg 1$ for the simple and double-threshold boxcar filters and $\Delta t/\tau > 8$ for the half-boxcar filter to achieve peak performance (see Section~\ref{sec:periodic:drop}).}
\label{tab:fingamma}
\end{table*}

\subsection{Initial drop in fidelity}\label{sec:periodic:drop}
 
The initial drop in fidelity $\Delta F_{\text{in}}$ for all three variants of the boxcar filter comes from single logical errors in the final averaging box that do not have time to be detected. There are two dominant types of logical error: a single parity misidentification, or a single bit flip that happens too late within the averaging period. Other errors are higher-order and comparatively negligible. Since there are two parities, a misidentification can occur with probability $2 P_{\text{mis}}$. The case of the bit flip requires more careful analysis, since it may occur at any point within the final box. 

For bits 1 and 3, if a flip happens at time $\delta t$ after the center of the last box, $\Delta t/2 + \delta t$, the probability of not detecting the flip is $P(\bar{r}>0\,|\,r_m)$ with a shifted signal mean of $r_m = 2\delta t/\Delta t$ and variance $\tau/\Delta t$ as in Eq.~\eqref{eq:halfboxerror}. There are two possibilities of a bit flip, so their contribution to the initial drop is
\begin{align}
 2\,\frac{\mu\Delta t}{2}\,\int_{-1}^1 P(\bar{r}>0\,|\,r_m)\,dr_m = \mu\Delta t.
\end{align}

To detect a bit 2 flip correctly, both integrated signals should be less than the threshold $a$ (where $a=0$ for boxcar and half-boxcar filters and $a > 0$ for the double-threshold filter). The negation of this is for one of the signals to be greater than $a$. To not double-count the errors for bits 1 and 3, the remaining signal should also remain greater than 0. Therefore, a flip is not detectable if $\bar{r}_{1,2}>0$ and $\bar{r}_{2,3}>a$, or vice versa. If we denote one such event $A$ and the reverse configuration $B$, then the total probability of not detecting the bit 2 flip is $P(A \cup B)=P(A)+P(B)-P(A \cap B)$, where the intersection $A \cap B$ has both signals greater than $a$. After summing this probability for all $r_m$ we obtain
\begin{align}\label{eq:p2}
  P_2 &\equiv \int_{-1}^1 [2P(\bar{r}>a\,|\,r_m)P(\bar{r}>0\,|\,r_m) \nonumber \\
  &\qquad\quad - P(\bar{r}>a\,|\,r_m)P(\bar{r}>a\,|\,r_m)]\,dr_m.
\end{align}
The total contribution of this scenario to the initial drop is thus, $(\mu\Delta t/2)P_2$.

Gathering the above contributions, the total initial drop in fidelity for all boxcar filters is
 \be
 \Delta F_{\text{in}} = \left[2 + P_2 \right]\frac{\mu\Delta t}{2} + 2\,P_{\text{mis}}.
 \ee
 
Since this formula involves an unwieldy integral, we will find suitable approximation before continuing. As will become clear in the next section, for the simple boxcar and double-threshold filters the averaging duration $\Delta t$ should be set fairly long compared to $\tau$ to achieve good performance. In this regime, we $P_2 \approx 1$ is an excellent approximation. Similarly, we will find that $P_{\text{mis}}$ is negligible for both the simple boxcar and double-threshold filters when $\Delta t \gg \tau$. 

For the half-boxcar filter, the peak performance will be achieved for significantly smaller $\Delta t$, so we must evaluate $P_2$ more precisely to obtain an accurate estimate. When $a=0$, $P_2$ acquires the asymptotic form
\begin{align}\label{eq:p2boxcar}
  P_2 &\xrightarrow{a=0,\,\Delta t > 8\tau} 1 - \frac{1}{\sqrt{\pi\Delta t/\tau}} + \frac{\exp(-\Delta t/\tau)}{(\Delta t/\tau)\sqrt{\pi}},
\end{align}
which converges very slowly to 1 as $\Delta t \to \infty$. For $\Delta t \sim 15\tau$, $P_2 \sim 0.85$. The dominant part of this asymptotic form must be kept, as well as the $P_{\text{mis}}$ contribution to the error rate. Anticipating these simplifications now and using the asymptotic formula $P_{\text{mis}} \approx \exp(-\Delta t/2\tau)/\sqrt{2\pi \Delta t/\tau}$ thus yields the final approximations appropriate for the parameter regimes that yield peak filter performance:
\begin{align}
	\label{eq:finboxcar}
 \Delta F_{\text{in}} &\xrightarrow{\text{boxcar}} \frac{3}{2}\mu\Delta t \\
	\label{eq:finhalfbox}
 \Delta F_{\text{in}} &\xrightarrow{\text{half-box}} \frac{3}{2}\mu\Delta t - \frac{\mu\sqrt{\tau\Delta t}}{2\sqrt{\pi}} + \frac{\sqrt{2}\, e^{-\Delta t/2\tau}}{\sqrt{\pi \Delta t/\tau}} \\
	\label{eq:findouble}
 \Delta F_{\text{in}} &\xrightarrow{\text{doub.thr.}} \frac{3}{2}\mu\Delta t
\end{align}
In Section~\ref{sec:simulation} we verify these expressions numerically for the bit-flip rates $\mu\tau \in [10^{-6},10^{-3}]$.

The final expressions for the logical error rate $\Gamma$ and initial fidelity drop $\Delta F_{\text{in}}$ are summarized in Table~\ref{tab:fingamma}.

\subsection{Optimizing filter parameters}\label{sec:periodic:boxcaropt}

The boxcar filters contain several tunable parameters that play the role of the prior information about $\mu$ and $\tau$ required for the Bayesian filter. That is, the averaging duration $\Delta t$ and asymmetric threshold $a$ must be set prior to processing the parity signals. To achieve peak filter performance, these parameters must be optimized for each $\mu$ and $\tau$. As such, in order to fairly compare the performance of the boxcar filters to the Bayesian filter, we must choose an appropriate optimization strategy for the boxcar filter parameters.

We choose to optimize the filter parameters to minimize the total decay in average fidelity in the linear decay regime. Both the initial fidelity drop $\Delta F_{\text{in}}$ and the linear decay rate $\Gamma$ contribute to the total infidelity, so we optimize both together by maximizing the duration of time required for the average fidelity to drop a total of 10\%. As discussed after Eq.~\eqref{eq:fidnotrack}, a 10\% fidelity drop is roughly the maximum tolerable drop while remaining in the linear decay regime without error correction, which makes it a reasonable target. Since $F(t) = 1 - \Delta F_{\text{in}} - \Gamma t$ for a duration $t$ in this linear regime, this optimization procedure yields a maximum time to drop by $10\%$ fidelity: 
\begin{align}\label{eq:maxt}
	t_{\text{max}} = \max_{\text{params}} \frac{0.1 - \Delta F_{\text{in}}}{\Gamma}.
\end{align}
For most cases, the duration $t_{\text{max}}$ will be sufficiently long that the linear decay dominates the total average-fidelity decay, making this procedure essentially equivalent to minimizing the decay rate $\Gamma$ directly. However, for some cases with larger flip rates $\mu$ the initial drop becomes too large to neglect and this maximum-time optimization produces more reasonable results. 

We now systematically optimize the general formulas for the initial drop in fidelity $\Delta F_{\text{in}}$ and the logical error rate $\Gamma$, using the maximum drop-time procedure outlined in Eq.~\eqref{eq:maxt}. The post-optimization formulas will show the best achievable performance of each filter more clearly. For sake of simple comparisons, we only keep the dominant scaling of each analytical approximation in what follows. However, we will numerically optimize the full formulas in Table~\ref{tab:fingamma} and use the full expressions when comparing theory to numerical simulations in Section~\ref{sec:simulation}.

\subsubsection{Boxcar filter}

The boxcar filter has only one free parameter to optimize: the boxcar duration $\Delta t$. We use the optimization procedure of Eq.~\eqref{eq:maxt}, with the logical error rate from Eq.~\eqref{eq:gammaboxcar} and initial drop from Eq.~\eqref{eq:finboxcar}. We use the error function approximation $P_\text{mis} \approx \exp(-\Delta t/2\tau)/\sqrt{\tau/2\pi\Delta t}$ from Eq.~\eqref{eq:pmis} to make the formulas analytically tractable. We solve for the minimum by taking a derivative of Eq.~\eqref{eq:maxt} with respect to $\Delta t$ and setting it to zero, in the usual way, which produces the results
\begin{align} 
\frac{\Delta t}{\tau} &\approx  0.207\, (\mu \tau)^{-2/3} - 1.3\,(\mu \tau)^{-1/3} + 6, \nonumber \\
\Gamma \tau &\approx 1.86\,(\mu \tau)^{4/3}, \\
\Delta F_{\text{in}} &\approx 0.31\,(\mu\tau)^{1/3}, \nonumber
\end{align}
where we have replaced purely numerical prefactors with decimal approximations and have truncated the expressions to remove negligible terms. We choose the precision of these numerical constants and the truncations so that the simplified analytical formulas closely reproduce the numerically optimized result. 

Notably, the optimal averaging duration $\Delta t$ dominantly scales as $\mu^{-2/3} \propto \Gamma^{-1/2}$, so becomes impractically long for small error-rates $\Gamma$. Moreover, since the logical error rate $\Gamma$ scales as $\mu^{4/3}$, the filter performs dramatically worse than the $\mu^2$ scaling of the Bayesian filter. These features make the simple boxcar filter ill-suited for practical error correction. 

\subsubsection{Half-boxcar filter}

To optimize the half-boxcar filter over the duration $\Delta t$, we follow the same procedure as in the previous section for the boxcar filter. We use Eqs.~\eqref{eq:gammahalfbox} and \eqref{eq:finhalfbox} and the error function approximation from Eq.~\eqref{eq:pmis}. The minimization procedure by taking a derivative with respect to $\Delta t$ produces the approximate nonlinear relation $\exp(\Delta t/2\tau) = 3/14\mu\tau\sqrt{\pi\Delta t/2\tau}$, which we solve recursively for $\Delta t/\tau$. This procedure yields the following continued fraction as a perturbative solution
\begin{align}\label{eq:halfboxdeltat}
	\frac{\Delta t}{\tau} &\approx 2\ln\frac{3}{14\mu\tau\sqrt{\pi\ln[3/(14\mu\tau\sqrt{\pi\,\cdots})]}} \\
	&\approx 2 \ln\frac{1}{15\mu\tau}, \nonumber
\end{align}
where the final approximation truncates the recursion at the dominant logarithmic functional form. The constant inside the logarithm is chosen as a crude fit to the full numerically optimized curve within the parameter regime $\mu\tau \in [10^{-6},10^{-3}]$ to help simplify the scaling comparison. Using this simplification, the logical error rate and initial drop have the forms
\begin{align} \label{eq:halfboxgamma} 
\Gamma \tau &\approx 8.4\,(\mu\tau)^2 \ln \frac{1}{15 \mu \tau}, \\
\Delta F_{\text{in}} &\approx 3(\mu\tau)\ln \frac{1}{15 \mu \tau}, \nonumber 
\end{align}
where numerical constants have again been reduced to appropriate precision decimals based on fits to the numerically optimized results.

Notably, the logical error rate $\Gamma$ now scales with $\mu^2$, up to logarithmic corrections, analogously to the Bayesian filter in Eq.~\eqref{eq:bayesiangamma}, making the half-boxcar filter suitable for practical error correction. In the full numerical simulations that we detail in the following section, we will see that of the boxcar-averaging filters the half-boxcar variation has the closest performance to the optimal Bayesian filter, despite its dramatic reduction in computational overhead.

\subsubsection{Double threshold filter}

\begin{figure}[t]
\includegraphics[width=0.9\columnwidth]{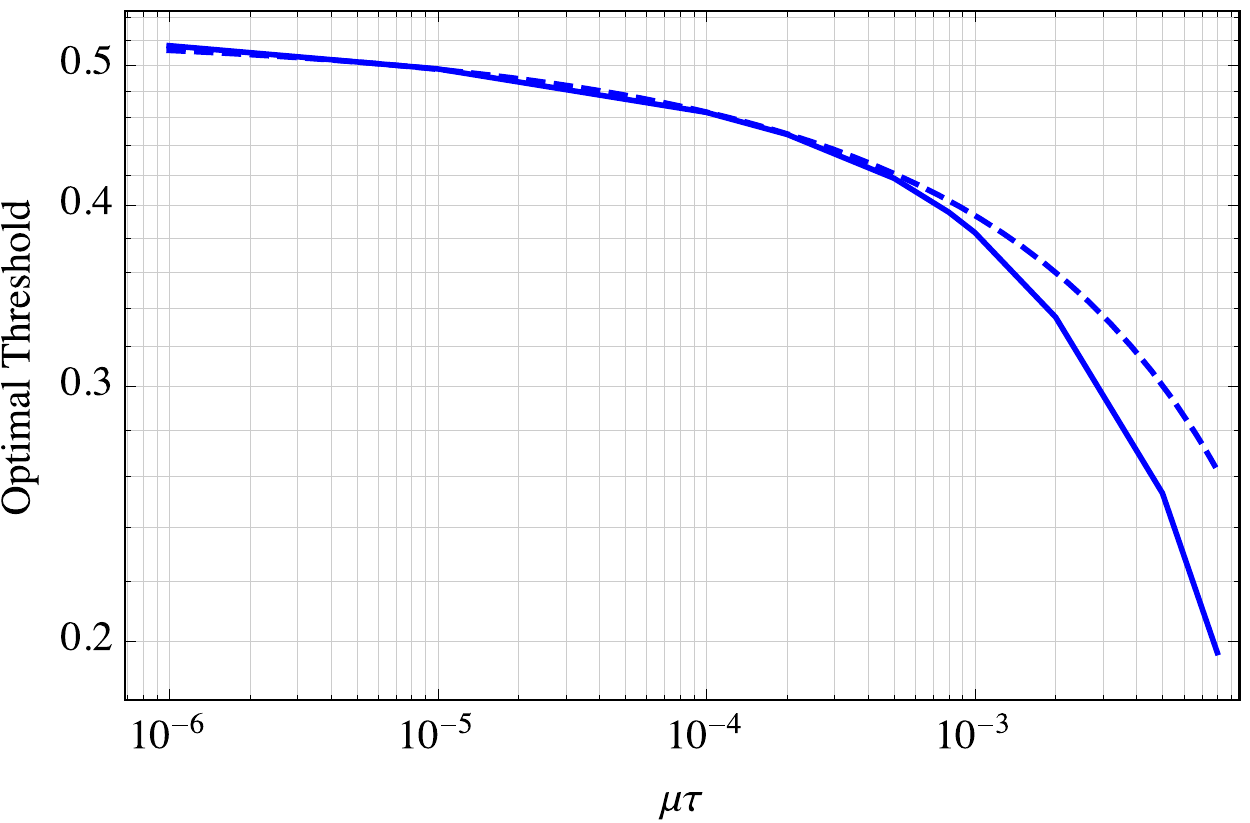}
\caption{Numerically optimized second threshold $a\geq 0$ for the double-threshold filter, as a function of the bit-flip rate $\mu\tau$. The dashed line is a crude analytical fit $a \approx 0.525[1 - 2.5(\mu\tau)^{1/3}]$ within the range $\mu\tau \in [10^{-6},10^{-3}]$.}
\label{fig:optimala}
\end{figure}

\begin{table*}[t]
\begin{center}
\setlength{\tabcolsep}{8pt}
\renewcommand{\arraystretch}{4}
\begin{tabular}{l|l|l|l|l}
\multicolumn{1}{c|}{\textbf{Filter}} & $\Delta F_{\textrm{in}}$ & $\Gamma \tau$ & $\Delta t/\tau$ & $a\geq 0$ \\
\hline\hline
Bayesian & $\displaystyle 1.5\,(\mu\tau)\ln\frac{1}{\mu\tau}$ & $\displaystyle 3\,(\mu\tau)^2\ln\frac{1}{\mu\tau}$ & -- & -- \\
\hline
Boxcar & $\displaystyle 0.31\,(\mu\tau)^{1/3}$ & $\displaystyle 1.86\,(\mu\tau)^{4/3} $ & $0.207\, (\mu\tau)^{-2/3}$ & $0$ \\
\hline
Half-boxcar & $\displaystyle 3\,(\mu\tau)\ln\frac{1}{15\mu\tau}$ & $\displaystyle 8.4\,(\mu\tau)^2\ln\frac{1}{15\mu\tau}$ & $\displaystyle 2\,\ln\frac{1}{15\mu\tau}$ & $0$ \\
\hline
Double-threshold & $\displaystyle 12\,(\mu\tau)\ln\frac{1}{150\mu\tau}$ & $\displaystyle 33\,(\mu\tau)^2\ln\frac{1}{150\mu\tau}$ & $\displaystyle 8\,\ln\frac{1}{150\mu\tau}$ & $0.5$ 
\end{tabular}
\end{center}
\caption{Dominant scaling with $\mu\tau$ of optimized filter performance. We compare the initial fidelity drops $\Delta F_{\text{in}}$, logical error rates $\Gamma$, boxcar averaging durations $\Delta t/\tau$, and second thresholds $a\geq 0$. We show only the dominant scaling in each case, truncating smaller corrections for sake of simple comparison. We choose the precision of numerical prefactors to best fit the full numerically optimized results obtained from Table~\ref{tab:fingamma} in the regime $\mu\tau \in [10^{-6},10^{-3}]$. See Section~\ref{sec:periodic:boxcaropt} for details on the optimization procedures.}
\label{tab:optscaling}
\end{table*}

Unlike the preceding boxcar filters, the double-threshold filter has two free parameters to optimize: the boxcar duration $\Delta t$, and the second threshold $a\geq 0$. To get a rough idea of the analytic scaling, we follow a crude sequential optimization strategy. First, we follow the same optimization procedure from the previous section for $\Delta t$ for a fixed $a$, using Eqs.~\eqref{eq:gammadouble} and \eqref{eq:findouble} and the error function approximation from Eq.~\eqref{eq:pmis}. This optimization again yields a continued fraction solution
\begin{align}\label{eq:thresholddeltat}
\frac{\Delta t}{\tau} &\approx \frac{2}{(1-a)^2} \ln\frac{(1-a)^2}{6 \mu\tau  \sqrt{\pi\ln\frac{(1-a)^2}{6 \mu\tau\sqrt{\pi\,\cdots }}}}, \\
&\approx \frac{2}{(1-a)^2} \ln\frac{(1-a)^2}{6 \mu\tau  \sqrt{\pi\ln\frac{(1-a)^2}{6 \mu\tau\sqrt{\pi}}}}, \nonumber 
\end{align}
where the final approximation truncates the recursion at the second-order, which achieves better accuracy than the first-order truncation in the parameter regime $\mu\tau \in [10^{-6},10^{-3}]$. We find numerically that the dependence of the optimal $a$ on $\mu\tau$ varies slowly in this parameter regime, so we approximate it using the following function
\begin{align}\label{eq:thresholda}
a &\approx 0.525[1 - 2.5(\mu\tau)^{1/3}], 
\end{align}
which is a crude fit to the numerically optimized result and not derived analytically. We compare this crude formula to the numerically optimized result in Fig.~\ref{fig:optimala} for completeness. We will use these approximate analytical fits to the numerical optimization in Figs.~\ref{fig:fidelity}--\ref{fig:optimalgamma} in the following section.

For the purposes of comparing the dominant scaling of the various filters, we further approximate the threshold as constant, $a\approx 0.5$, and truncate the recursive solution of $\Delta t/\tau$ to first-order, which yields the following loose approximations to the logical error rate and initial drop in fidelity,
\begin{align} \label{eq:thresholdgamma}
\frac{\Delta t}{\tau} &\approx 8\ln\frac{1}{150\mu\tau}, \nonumber \\
\Gamma\tau &\approx  33\, (\mu\tau)^2 \ln \frac{1}{150 \mu \tau}, \\
\Delta F_{\text{in}} &\approx  12\,(\mu\tau)\ln \frac{1}{150 \mu \tau}. \nonumber 
\end{align}
As with the preceding filters, we have replaced constants factors with to best fit the full numerical optimization. The intention of these final formulas is not to be exceptionally accurate, but rather to capture the crude dominant scaling for sake of simple comparison with the other filters.

As with the half-boxcar filter, the double-threshold filter achieves the $\mu^2$ scaling of the logical error rate $\Gamma$, up to logarithmic corrections, so is also a suitable filter for error correction. It has the benefit of being Markovian, as opposed to the half-boxcar filter that requires memory, but its scaling prefactors are not as favorable as in Eq.~\eqref{eq:halfboxgamma} for the half-boxcar filter or Eq.~\eqref{eq:bayesiangamma} for the Bayesian filter. For convenience, we compare the dominant scaling of all filters in Table~\ref{tab:optscaling}.

\section{Numerical Simulations}\label{sec:simulation}

\begin{figure*}[ht!]
\begin{center}
\subfloat[]{\includegraphics[width=0.45\textwidth]{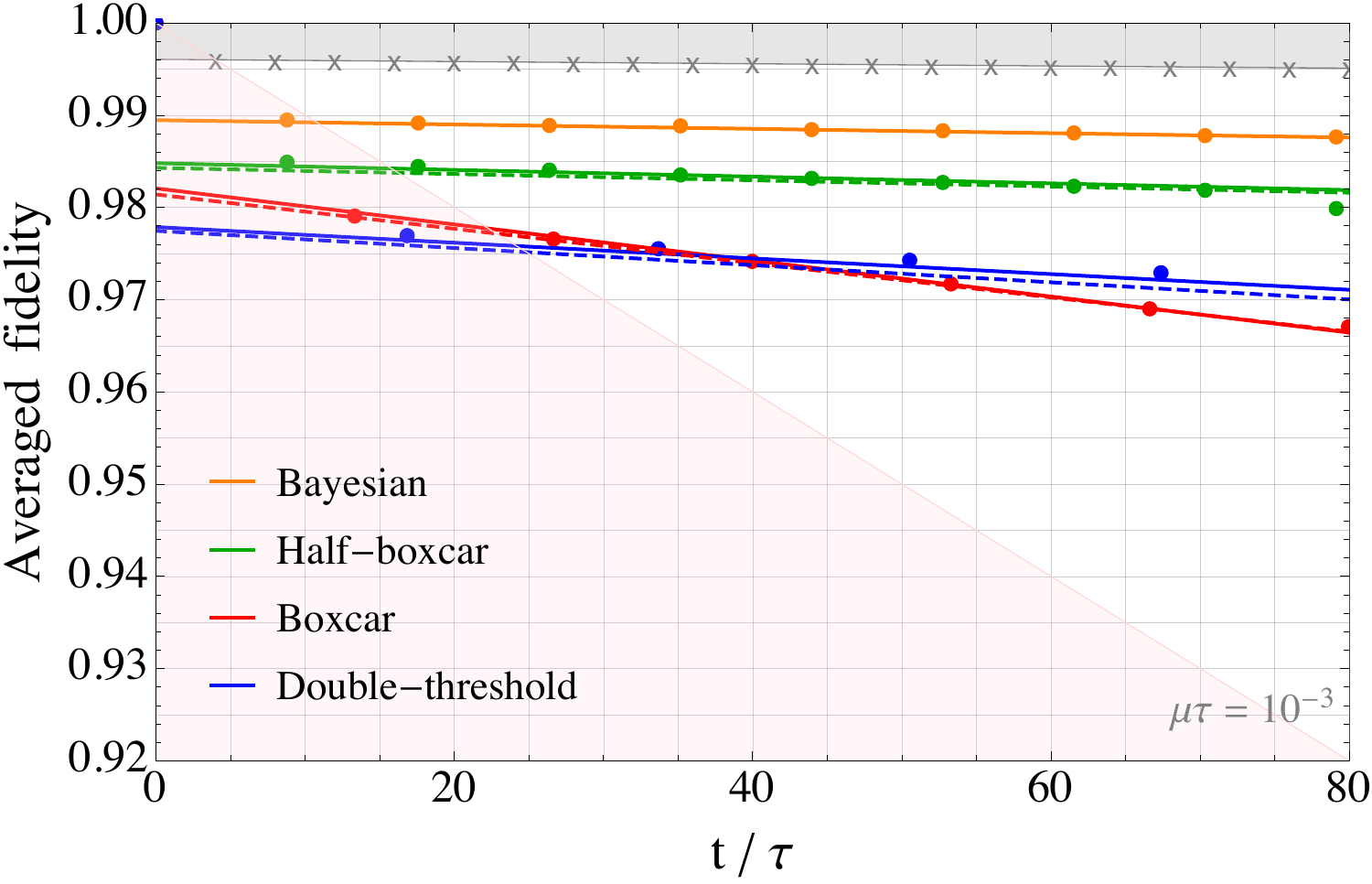}
	\label{fig:fidelity}} \quad
\subfloat[]{\includegraphics[width=0.45\textwidth]{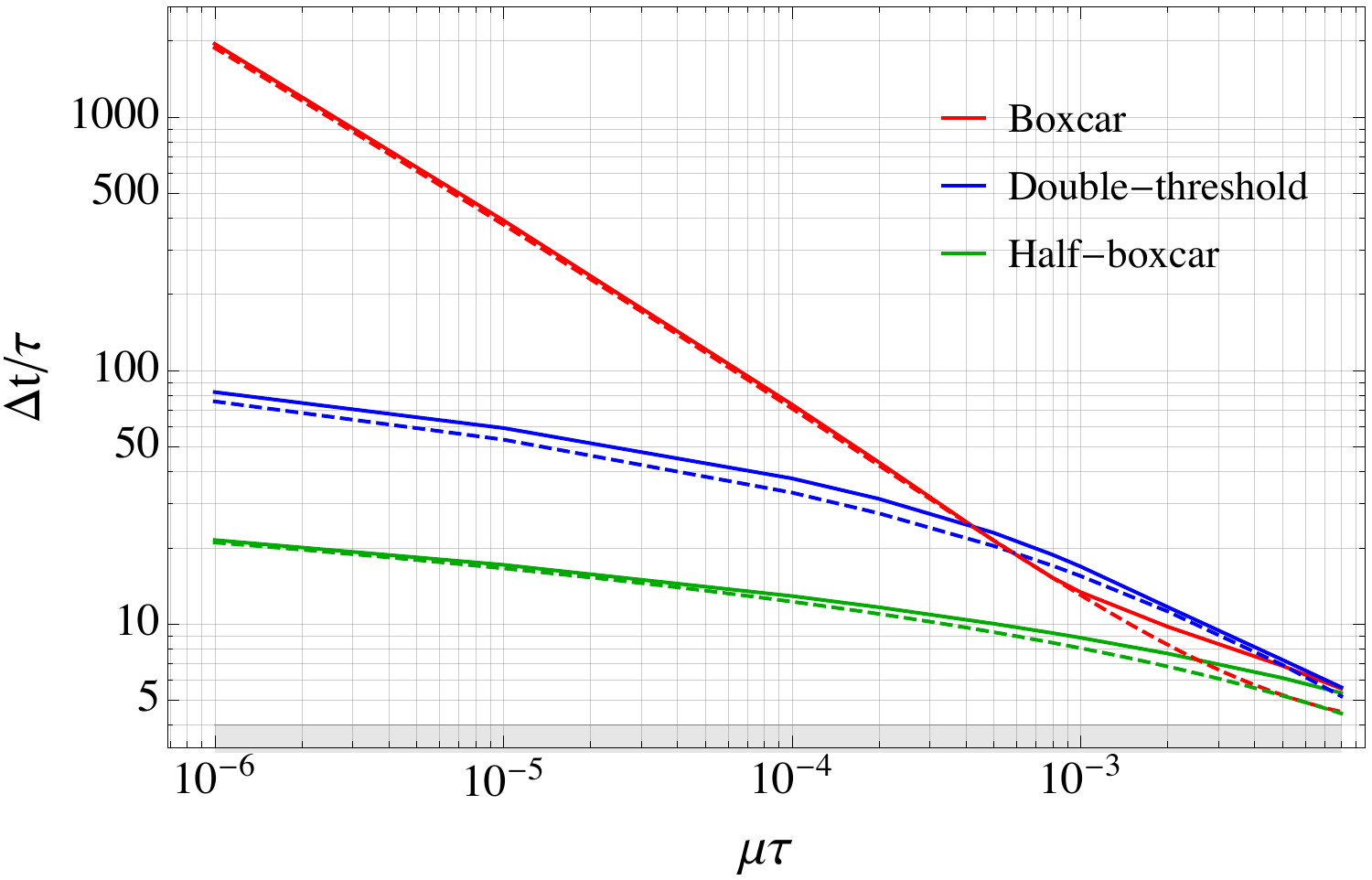}
	\label{fig:optimalboxcar}} \\
\subfloat[]{\includegraphics[width=0.45\textwidth]{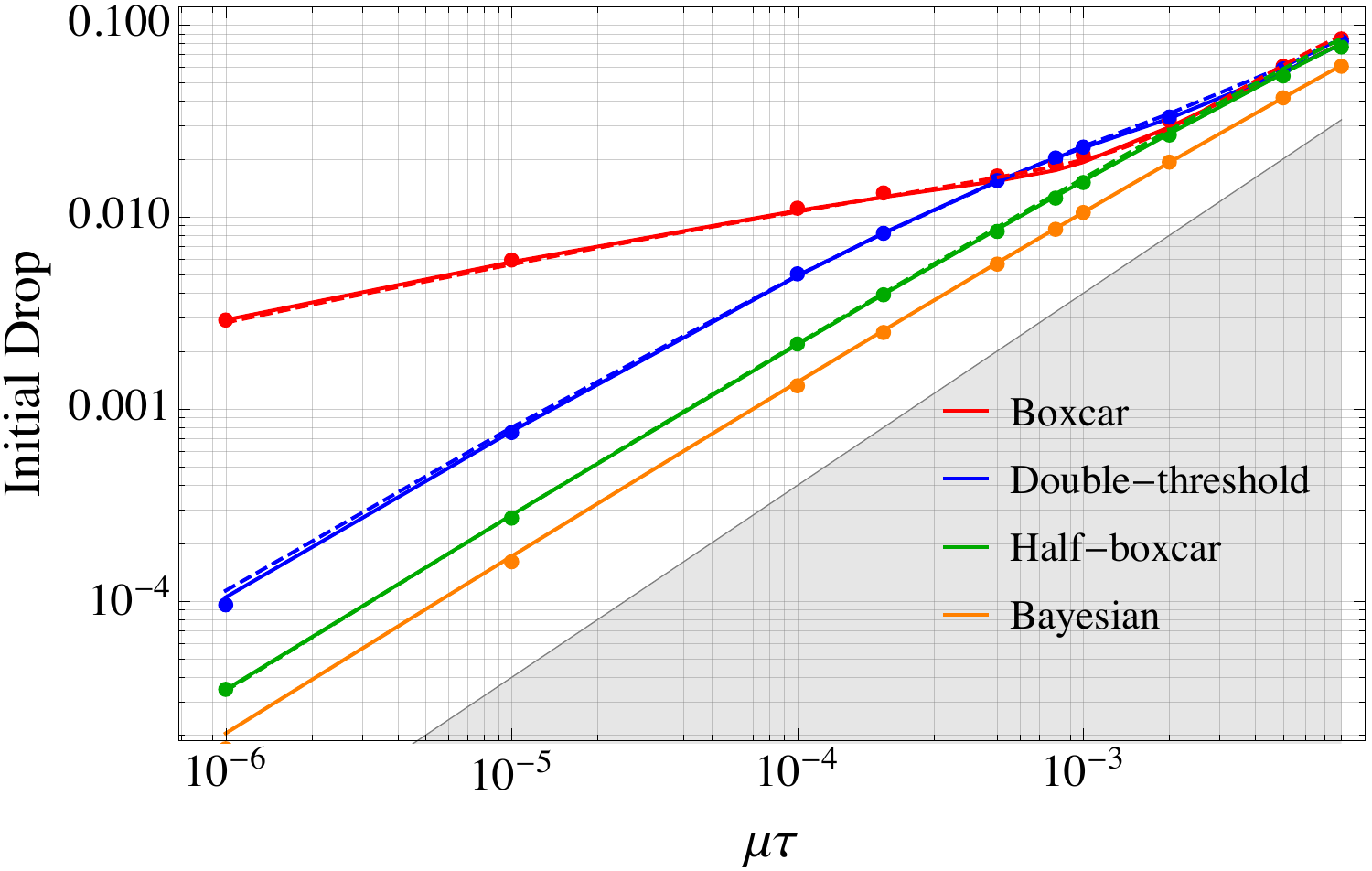}
	\label{fig:optimaldrop}} \quad
\subfloat[]{\includegraphics[width=0.45\textwidth]{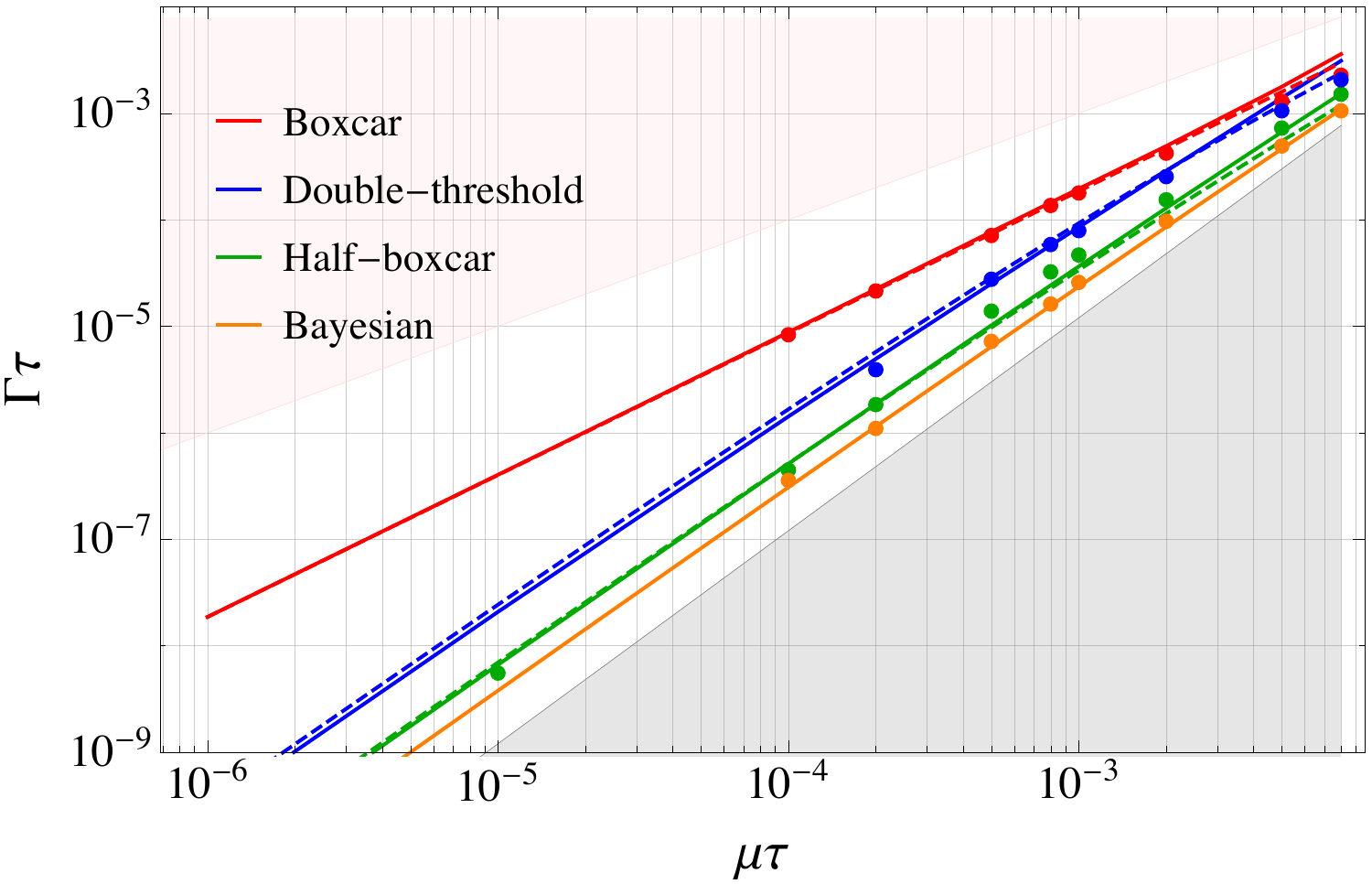}
	\label{fig:optimalgamma}} 
\end{center}
	\caption{Comparison of numerical simulations to analytical expressions. (a) Average fidelity $F(t)$ in time, optimized for a bit-flip rate of $\mu\tau = 10^{-3}$, showing how the initial drop $\Delta F_{\text{in}}$ and logical error rate $\Gamma$ of various error correction methods manifest. For the boxcar filters the data points indicate the box periodicity $\Delta t$, while data points on the Bayesian curve are sampled similarly for reference. Note that the last box of the half-box filter decays with the same rate as the basic boxcar filter, since the non-Markovian correction cannot be applied. (b) Optimized boxcar-averaging duration, $\Delta t/\tau$, with the time-continuous Bayesian case omitted. (c) Optimized initial drop in average fidelity $\Delta F_{\text{in}}$ for different filters, as a function of the bit-flip rate $\mu\tau$. (d) Optimized logical error rate $\Gamma$ for different filters, as a function of the bit-flip rate $\mu\tau$. The order in the legend from top-to-bottom matches the order of the curves on the left edge of each plot. In all plots, numerical simulations of the various filters using optimized parameters are plotted as circular data points with error bars smaller than the width of the points. The numerically-optimized analytical formulas in Table~\ref{tab:fingamma} are plotted as solid lines, while the simplified formulas in the main text are plotted as dashed lines. The shaded gray areas indicate the regions that are unattainable even by an ideal 3-qubit code, while the shaded red areas indicate the regions that are worse than that of a single qubit with no error correction. The analytical initial drops for the Bayesian filter include the prefactor correction $5/4\mapsto 3/2$ discussed in Section~\ref{sec:continuous:analysis}. The analytical initial drops for the boxcar filters numerically correspond to the time at half of the first box $\Delta t/2$. 
}
\end{figure*}

To check the validity of our filter analysis, we numerically implement continuous measurements of the 3-bit code, the linear Bayesian filter, and the three boxcar filter variations in the programming language \texttt{julia} \cite{tomasi2018}. To do this efficiently, we first pick a target bit-flip rate $\mu$ to test, such that $\mu\tau \in [10^{-6},10^{-3}]$ with $\tau$ being the reference timescale for the numerics (set to 1 for convenience). We then initialize a $3\times N$ array of bits to describe $N=\mathtt{floor}(10\,\text{max}\Delta t/dt)$ time steps of duration $dt = \tau/10$. The timescale $\text{max}\Delta t$ is the maximum optimal boxcar size of the four filters being tested for each $\mu$, and ensures that enough data is simulated per trajectory to assess the behavior of each filter. We compared the results to those obtained with $dt = \tau/100$ to verify that no residual time discretization artifacts were present in the numerics. The optimal box sizes are numerically determined from the formulas presented in Table~\ref{tab:fingamma}, and shown in Fig.~\ref{fig:optimalboxcar}, with maxima ranging from $\text{max}\Delta t/\tau \in [5,2000]$. 

To simulate each trajectory, at the initial time the bit state is set to $000$, representing the encoding $III$. Random bit flips are then added with Poisson statistics at the rate $\mu$. Specifically, the wait-time distribution for $n$ steps between two successive jumps is exponential, $p(n) = \exp[-n/(\mu dt)]/(\mu dt)$; for each qubit we sample this wait-time distribution to find the random number of steps $\mathtt{floor}(n)$ until the next jump for each bit, then flip the appropriate bits between specified jumps. After this procedure the $3\times N$ array holds the ``true'' state trajectory for the 3-bit code. This numerical model is thus a direct implementation of the hidden Markov model in Fig.~\ref{fig:Markov} that is described in the main text. 

Given a true 3-bit trajectory, we then simulate the noisy parity signals $r_{i,j}$ by computing the exclusive-or $x(i,j)$ between neighboring bits $i$ and $j$ at each timestep $dt$, then constructing $r_{i,j} = -2x_{i,j} + \xi$, where $\xi$ is sampled from a normal distribution with mean $+1$ and variance $\tau/dt$. This construction vectorizes the noise simulation efficiently, and centers the mean signals for even parities at $+1$ and odd parities at $-1$. The resulting noisy signals then simulate the parity signals that one would obtain from performing continuous direct parity measurements in the laboratory, after the signals have been correctly normalized. 

Given the simulated noisy signals $r_{i,j}$, we then pass both signals through each of the four trial filters analyzed in the preceding section: linear Bayesian, simple boxcar, half-boxcar, and double-threshold boxcar. We set the tunable parameters, $\Delta t$ and $a$, for the three boxcar filters to optimal values determined by the numerical optimization of the formulas in Table~\ref{tab:fingamma}. (We also verify numerically that tuning these parameters away from the theoretical optimum correctly shows that the parameter values are optimum.) Each filter then returns a $3\times N$ array of estimated 3-bit state trajectories. For each $dt$, we compute the bit state fidelity as a simple equality test between the triplet of true bits and the triplet of estimated bits, yielding 1 if the bits agree and 0 if they disagree. We compute the average state fidelities by repeating this process between $10^6$ and $10^8$ times and averaging the fidelities at each time step. 

This simulation procedure produces the numerical results plotted as the points in Figs.~\ref{fig:fidelity}--\ref{fig:optimalgamma}, with final numerical error bars on the order of the width of the points or smaller. The solid lines show the formulas summarized in Table~\ref{tab:fingamma} after numerical parameter optimization. For reference, the dashed lines show the crude analytical approximations of the optimized formulas that we presented in Section~\ref{sec:periodic:boxcaropt}. (Note that for the double-threshold filter in Figs.~\ref{fig:optimalboxcar}--\ref{fig:optimalgamma} we plot the more accurate $a$-dependent analytic formulas in Eqs.~\eqref{eq:thresholddeltat} as the dashed lines.) For the boxcar filters, we found that in order to apply the linear decay formula in Eq.~\eqref{eq:lineardecay} to the simulated data, the initial drop $\Delta F_{\text{in}}$ should be placed in the middle of the first averaging box, at $t=\Delta t/2$, after which the linear fit with slope $\Gamma$ correctly describes the data. For smaller $\mu$ the logical error rates $\Gamma$ become quite small so require more realizations to resolve the average to sufficient numerical precision; in the cases of the double-threshold and simple boxcar filters the optimized durations $\Delta t/\tau$ were sufficiently long to prohibit accurate averaging of $\Gamma$ below $\mu\tau \sim 10^{-4}$. Nevertheless, for all successfully simulated results the agreement is excellent between numerical simulations and numerically optimized analytical formulas from Table~\ref{tab:fingamma}. 

In Fig.~\ref{fig:fidelity}, we show the time-dependent average fidelities $F(t)$ of all methods, optimized for a relatively large bit-flip rate of $\mu\tau = 10^{-3}$. The numerical simulations (data points) confirm the numerically-optimized analytical results for $\Delta F_{\text{in}}$ and $\Gamma$ summarized in Table~\ref{tab:fingamma} (solid lines), as well as the corresponding crude approximations (dashed lines). The gray line is a simulation of idealized 3-bit code error correction using perfect-fidelity projective parity measurements with a rapid cycle delay of $\delta t/\tau = 4$ (see the Appendix for details); the shaded gray region above this line roughly represents fidelities that are inaccessible to even an ideal implementation of the 3-bit code. The light red shaded region below indicates fidelities that are worse than that of a single bit without error correction. 

After the initial drops in fidelity $\Delta F_{\text{in}}$ in Fig.~\ref{fig:fidelity}, the Bayesian filter (orange, top curve) and half-boxcar filter (green, second curve from top) achieve asymptotic slopes (corresponding to the logical error rates $\Gamma$) that are comparable to that expected for ideal operation of the code. The double threshold filter (blue, bottom curve at left of graph) performs slightly less favorably, while the simple boxcar filter (red, third curve from top at left of graph) performs significantly worse, as anticipated in the previous section. For the half-boxcar filter, the last box decays at the same rate as the simple boxcar filter because the non-Markovian correction cannot be applied to the last box. This change in decay rate in the final box yields an additional contribution to the net fidelity drop $\Delta F_{\text{fin}} \equiv \Delta t(\Gamma_{\text{boxcar}}-\Gamma_{\text{half-boxcar}})$ that is not observed with the Markovian filters. 

In Fig.~\ref{fig:optimalboxcar}, we show the optimized averaging durations $\Delta t$ for the boxcar methods (solid lines) and corresponding crude approximations (dashed lines). The gray line at bottom indicates the rapid cycle delay of $4\tau$ for the idealized projective measurements. The non-Markovian half-boxcar filter achieves the shortest averaging durations with $\Delta t \lesssim 20\tau$ even for small bit-flip rates of $\mu\tau \sim 10^{-6}$. The simple boxcar filter requires excessively long optimal averaging durations (up to two orders of magnitude longer than the half-boxcar filter for small bit flip rates). The Markovian double threshold filter consistently requires averaging lengths that are a factor of roughly 2--4 longer than the half-boxcar to achieve similar performance. 

In Fig.~\ref{fig:optimaldrop}, we show the optimized scaling of the initial drop $\Delta F_{\text{in}}$ with $\mu$ for all methods, using the same color and line-style conventions as Fig.~\ref{fig:fidelity}. The numerical simulations confirm the numerically-optimized analytical results in Table~\ref{tab:fingamma} for the entire tested range of parameters $\mu\tau \in [10^{-6},10^{-2}]$. For the Bayesian filter analytical curve we adjust the derived formula by making the substitution $5/4 \mapsto 3/2$ in Eq.~\eqref{eq:bayesiandrop}, which corrects a systematic deviation caused by noise-fluctuations, as discussed at the end of Section~\ref{sec:continuous:analysis}. Notably, the half-boxcar filter achieves an initial drop roughly a factor of 2 larger than the optimal Bayesian filter. For contrast, the double-threshold filter has an initial drop that is roughly a factor of 8 larger than the Bayesian filter. The simple boxcar filter suffers from comparatively large drops greater than $0.3$\%, even with small bit-flip rates of $\mu\tau \sim 10^{-6}$, which is nearly two orders of magnitude larger than the Bayesian filter.

In Fig.~\ref{fig:optimalgamma}, we show the optimized scaling of the logical error rate $\Gamma$ with $\mu$ for all methods, using the same conventions. The numerical simulations again confirm the numerically-optimized analytic results, up to small deviations for the half-boxcar filter at larger bit-flip rates. We anticipated this slight deviation between the analytics and the simulated data points for $\mu\tau > 10^{-4}$ during the derivation in Section~\ref{sec:periodic:halfbox}, where it arises from a short boxcar duration, $\Delta t \lesssim 15\tau$, that prevents convergence to the asymptotic behavior assumed in the analytical formulas. For the double-threshold and simple boxcar filters we simulate only larger bit-flip rates $\mu\tau \gtrsim 10^{-4}$ due to the optimal boxcar sizes $\Delta t$ becoming prohibitively long for smaller $\mu\tau$; however, the tested cases confirm the $\mu\tau$-dependence expected from the analytics.

These simulations confirm that quantum error correction based on \emph{passive state tracking} with continuous parity measurements is a viable strategy. As anticipated, the linear Bayesian filter performs the best, achieving only a slight reduction in performance compared to the idealized 3-bit code due to the noise of the monitored signal. Moreover, the half-boxcar filter nearly matches the Bayesian filter in performance despite a dramatic reduction in processing requirements, which makes it the best balance between performance and practicality of the minimal filters considered here. The double-threshold filter also scales comparably, though performs slightly worse overall. We also emphasize that in the presence of experimental nonidealities, realistic implementations of the 3-bit code that use entangling gates, ancillas, and projective measurements are likely to perform comparably to the continuous measurement filters considered here; for completeness, we provide a similar analysis of the ancilla-based projective case in the Appendix.

\section{Conclusions}\label{sec:conclusions}

We have analyzed the 3-qubit bit-flip code to assess the performance of direct methods for measuring the syndromes using time-continuous parity measurements. For interpreting the time-continuous noisy signals of the direct syndrome measurements, we have introduced and analyzed four distinct filters: (i) an efficient linear variation of an optimal Bayesian filter, (ii) a simple boxcar-averaging filter, (iii) a minimal non-Markovian ``half-boxcar'' variation of the boxcar-averaging filter, and (iv) a minimal Markovian variation of the boxcar-averaging filter that uses two thresholds. We have derived analytic estimations for the performance of all filters and have verified them with numerical simulations. 

These direct parity-measurement methods benefit from a reduction in hardware resources compared to ancilla-based methods (namely two fewer ancillary qubits), which limits the number of inherent bit-flip-error pathways even before extending the bit-flip code to more sophisticated encoding schemes. The Bayesian filter most closely approaches the ideal performance of the ancilla-based bit-flip code, but also requires the most computational resources for real-time processing of the noisy syndrome measurements. The boxcar variations require less active processing than the optimal Bayesian filter, so should be more easily implemented with signal processing hardware, such as field-programmable gate arrays (FPGAs), for the purposes of real-time syndrome tracking. The non-Markovian half-boxcar filter achieves the best balance between performance and computational overhead of the considered methods. The Markovian double-threshold filter performs slightly less well than the half-boxcar filter, but avoids the additional memory overhead at the expense of an increased boxcar duration. All three methods are suitable for immediate implementation with current superconducting hardware.

The results of our study are promising for the continued investigation of direct syndrome-measurement methods. However, three scalability issues that we have ignored need to be addressed before direct methods can achieve full quantum error correction. First, we have focused our analysis on the performance of the methods with respect to their intended design: protecting against bit-flip errors. As such, we have ignored other sources of infidelity, particularly dephasing of the parity subspaces due to imperfect overlap of the entangled microwave fields, which is analogous to ignoring entangling gate infidelity in analyses of ancilla-based error correction. Some analysis of these types of implementation imperfections has begun in recent years \cite{Criger2016,Ciani2017,Huembeli2017,Royer2018}, but more investigation is needed for a definitive assessment. Second, while we have presented a practical method for directly measuring the $ZZ$ parities needed for bit-flip correction, we have not addressed how to directly measure the $XX$ parities needed for additional phase-flip correction. Obtaining high-fidelity direct parity measurements for both $ZZ$ and $XX$ is an open problem currently under investigation. Third, high-fidelity extensions of direct two-qubit parity measurements need to be developed to implement more sophisticated error-correction schemes, such as the surface code that requires four-qubit parity measurements. 

A direct quantitative comparison of the performance of this continuous error correction to conventional implementations of gate-based ancilla plus projective-measurement is challenging. This is because different assumptions must be made about how the ancilla-based scheme is implemented, and what a fair comparison of the approaches is. In superconducting-based architectures, projective measurements have traditionally been implemented as thresholded continuous measurements anyway. Consequently, the always-on methods with a fast measurement rate have the obvious advantages of not needing time to implement the two-qubit gates, or to have any down time between repeating the cycle again, where other errors might sneak in. There is also the possibility of errors occurring in the ancilla qubits, which would then demand a much larger quantum circuit to make everything fault tolerant, but at the price of even more hardware. We analyze several different error scenarios that can occur in the gate-based implementation in the Appendices, for contrast. Our overall conclusion is that the hardware efficiency of the measurement-based parity has the potential to minimize error possibilities, assuming both good parity measurement fidelity and good gate fidelity.

Although full error correction using continuous parity measurements requires additional investigation, several experimental tasks can be achieved in the short term. First, the 3-qubit bit-flip code as analyzed here can be implemented immediately with current superconducting architectures. Second, a simple extension of the parity-syndrome monitoring idea to a 4-qubit Bacon-Shor error-detection code is a natural next step. Such a code involves four qubits in a square grid, coupled pairwise to parity-measuring resonators analogously to Fig.~\ref{fig:Experimental-setup}. 
A detailed analysis of the simultaneous measurement of $ZZ$ and $XX$ parities on the square grid is considered in Ref.~\cite{atalaya2017bacon}, which demonstrates that the error detection scheme works in a time-continuous way. A simpler variation of this idea can be performed without direct $XX$ measurements by alternating $ZZ$ measurements of different pairs, and suitably interleaving single-qubit rotation gates to effectively switch between $ZZ$ and $XX$ measurements. Such a variation is in between the usual ancilla-based projective scheme and a fully continuous scheme, much like the boxcar filters in the present work are in between ancilla-based schemes and fully continuous schemes. We expect such an experiment to be performed in the near future.

\begin{acknowledgments}
The authors thank Alexander N. Korotkov for many detailed discussions about the Bayesian and boxcar filters, in which he provided insight into the primary error mechanisms as well as initial derivations for the initial fidelity drops and average logical error rates. We also thank him for providing considerable feedback on the writing of this manuscript. The authors are additionally grateful for helpful discussions with Andrew Eddins, Machiel Blok, Leigh Martin, and William Livingston, as well as Liang Jiang. This work was supported by the Army Research Office (ARO) grant Nos. W911NF-15-1-0496 and W911NF-18-1-0178. JD also thanks Franco Nori for his hospitality at RIKEN during the completion of this manuscript, but does not thank the COVID-19 pandemic for subsequently delaying its publication.
\end{acknowledgments}



\appendix

\section{Ancilla-based parity measurements}
\label{app:ancilla}

\begin{figure}[b]
\begin{center}
\includegraphics[scale=0.45]{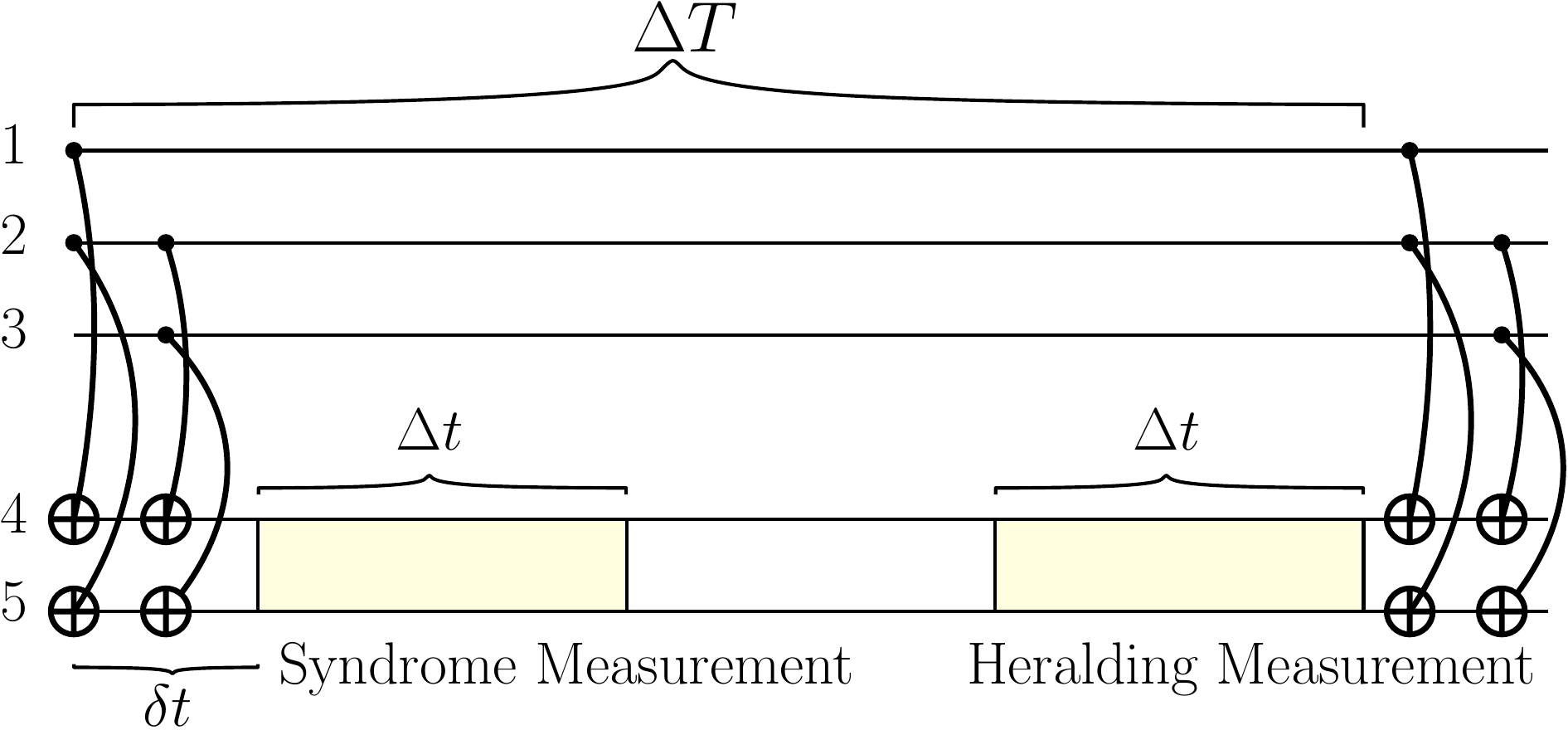}
\caption{
Ancilla-based error correction circuit for the 3-bit code. Successive controlled-NOT (CNOT) gates entangle two ancillas to the parities of bits 1-2 and 2-3 over a total gate time of $\delta t \sim 4\tau$. To measure the syndromes, the ancilla qubits are read out dispersively by integrating noisy signals for a duration $\Delta t$. After an arbitrary waiting time, the ancillas are reinitialized with a heralding measurement, also of duration $\Delta t$, which completes one full cycle time of $\Delta T$. For $\Delta T - \delta t < 2\Delta t$, the ancilla measurements fill the waiting time, so syndrome measurement and preparation heralding coincide.
}
\label{fig:circuit}
\end{center}
\end{figure}

Traditional ancilla-based error-correction with the three-bit code uses the correction circuit shown in Fig.~\ref{fig:circuit}, repeated over many cycles. At the beginning of the first cycle, we assume that the initial states of the ancillary qubits are known to be $|b_4\ra$ and $|b_5\ra$ with $b_4,b_5 \in \{0,1\}$. We couple the parity of qubits 1 and 2 ($P_{1,2}$) to ancillary qubit 4, as well as the parity of qubits 2 and 3 ($P_{2,3}$) to ancillary qubit 5, using a sequence of CNOT gates. For timing efficiency, the CNOTs may be performed in parallel so that the total duration of the gate sequence is the length of two CNOTs.

In the analysis that follows, we notate a CNOT that flips ancilla qubit $j$ conditioned on the excited state of qubit $k$ as $C_{j|k}$. A CNOT gate must be implemented by a sequence of more basic rotation gates that are hardware-dependent. For specificity, we model the CNOT with the \emph{cross-resonance gate} used by the IBM group \cite{Chow2011,Sheldon2016,magesan2020effective}. The cross-resonance interaction occurs when the data qubit $k$ is pumped at the resonance frequency of the ancilla qubit $j$, which can be modeled crudely by an effective interaction Hamiltonian $H \propto J(Z_kX_j + mI_kX_j)$ with a coupling strength $J$ and a chip-dependent cross-talk strength $m$ \cite{Chow2011} that is measurable \cite{Sheldon2016}. (For a more complete treatment of the cross-resonance gate, see Ref.~\cite{Korotkov2019}.) For a suitably timed pulse, a rotation phase of $\pm\pi/2$ can be accumulated by the $ZX$ interaction to yield the unitary gate: $\exp(-i (Z_k X_j)\pi/4 -i (I_k X_j)m\pi/4 )$. The simple cross-resonance interaction can take ${\sim 300}$--$400$ ns to complete \cite{Chow2011}, while echo optimizations can reduce the time to ${\sim 160}$ ns \cite{Sheldon2016}. After correcting the residual cross-talk rotation phase, the interaction produces a $(ZX)_{90}$ gate, which can be converted into a CNOT using single-qubit rotations and a phase correction, e.g.
\begin{align}\label{eq:cnot}
 C_{j|k} &= (I_k I_j)_{90}[(Z_k)_{-90}\otimes (X_j)_{90}](Z_kX_j)_{90} \\ 
  &= e^{-i I_k\otimes I_j\,\pi/4} (e^{i Z_k\pi/4}\otimes e^{i X_j\pi/4})e^{-i Z_k\otimes X_j\,\pi/4} \nonumber \\
  &= |0\ra\la0|_k\otimes I_j + |1\ra\la1|_k\otimes X_j. \nonumber
\end{align}
The single qubit gates take roughly ${\sim 30}$--$50$ ns to implement \cite{Corcoles2015}; therefore, we estimate the total CNOT gate to have an optimistic duration of $200$ ns in the following analysis, implying that the 2-CNOT gate time is $\delta t \sim 4\tau$ assuming a typical measurement timescale of $\tau \sim 100$ ns. For most purposes, it will be sufficient to use this final standard form of the CNOT. However, for analysis of potential logical error sources the specific gate implementation becomes important. 

We then measure $Z_4$ and $Z_5$ for the two ancillary qubits, using dispersive microwave readout similar to that outlined in Section~\ref{sec:setup}, and integrate the resulting signals for a duration $\Delta t$. After thresholding the integrated signals with a symmetric threshold of $a=0$, we obtain the pair of binary results $(R_4,\,R_5)$ with $R_4,R_5\in\{-1,1\}$. Using the knowledge of the initial ancilla states, we convert these results to the parity eigenvalues of $(Z_1Z_2,\,Z_2Z_3)$ with the relation $((-1)^{b_4}\,R_4,\,(-1)^{b_5}\,R_5)$. These eigenvalues are the error syndrome outcomes for tracking single bit flip errors with the same table used in Eq.~\eqref{eq:syndromes}.

We start the next cycle after a total cycle duration $\Delta T$, which includes the gate time $\delta t$, the syndrome measurement integration time $\Delta t$, an arbitrary waiting duration, and a heralding measurement time $\Delta t$. This heralding measurement is made to ensure the ancillas are in a known state before the next gate cycle, since bit flips may have occurred on the ancillas during the waiting time. If the waiting time is sufficiently short, such that $\Delta T - \delta t < 2\Delta t$, the readout and heralding measurements may be combined into a single long measurement of the ancillas. The only reason for separating the two for long cycle times is that long integration times may hide bit flips of the ancillas just prior to the start of the next cycle. If we assume that bit flips of the ancillas prior to the heralding measurement can be corrected, then there is no penalty for the measurements continuing for the entire waiting duration before thresholding. 

We use three strategies to estimate the performance of the ancilla-based case:
\begin{enumerate}
 \item \emph{Idealistic strategy} : We assume that the only errors which are not correctable are from higher-order data-qubit bit-flips. We also assume that the ancillas are not reused, so do not limit the cycle time. This is the optimal theoretical performance of the bit-flip code if no additional errors are introduced in the implementation. 
 \item \emph{Pessimistic strategy} : We keep all errors that arise from sudden bit flips of either the data qubits or the ancillas, including during measurement times and during the cross-resonance gate interaction. We use this strategy as a worst-case-scenario of the ancilla-based case that would occur without more sophisticated correction methods.
 \item \emph{Optimistic strategy} : We assume that all bit-flip-induced phase-flip errors can be avoided by embedding the 3-qubit code into a suitable fault-tolerant code, leaving only second-order errors. We assume non-Markovian processing of the syndrome history to remove the detectable ancilla-flip errors. We also assume that the ancillas are reused in each cycle. We use this strategy as a more realistic benchmark to assess the relative performance of the continuous case.
\end{enumerate}

The idealistic case acts as an effective lower bound for the infidelity to show how the realistic cases compare against the best possible theoretical case. In Figs.~\ref{fig:fidelity}, \ref{fig:optimaldrop}, and \ref{fig:optimalgamma} we plot this idealistic case as the gray line to provide a crude comparison for the continuous measurement cases. The pessimistic case acts as a worst-case scenario to show how the raw 3-qubit code would behave if implemented in a laboratory that only caused pure bit flips, with no sophisticated non-Markovian processing of the syndrome history. Although better processing techniques would certainly be employed to avoid most of these errors, the pessimistic case is still a useful comparison as a lower bound to bracket expectations. The optimistic case acts as a more realistic in-between scenario that incorporates plausibly achievable corrections. For evaluating realistic QEC, additional enhancements beyond the simple bit-flip code should be assumed, which makes the optimistic strategy a reasonable performance comparison. 

We now derive expressions for the initial drop in fidelity $\Delta F_{\text{in}}$ and the logical error rate $\Gamma$ for these projective measurement scenarios, to be compared to the continuous measurement case. In each case, we fix the gate time $\delta t = 4\tau$ to the length of two CNOT gates and optimize over two free filter parameters: measurement integration time $\Delta t$ and cycle repetition time $\Delta T \geq \Delta t + \delta t$. As a caution, when experimentally implemented the value of the measurement timescale $\tau$ will differ between the continuous parity measurements of the main text and the ancilla measurements in the appendix. Parity measurements are likely to have a timescale $\tau$ that is several times longer than the $\tau$ achievable for the measurement of a single qubit. For simplicity of notation, however, we ignore this distinction in what follows; nevertheless, care should be taken when comparing results to the continuous measurement cases.

\subsection{Logical error rate: Idealistic case}\label{app:ancilla:ideal}

In the idealistic case, we ignore all errors introduced by the ancilla qubits, ignore all errors that could happen during CNOT gates, ignore phase flips, and ignore misidentification errors of the readout. We also assume that all ancilla measurements can be done in parallel with an infinite supply of ancillary qubits, so that there is no restriction on repetition time arising from reuse of the ancillas. As such, the cycle time is limited only by the gate time, $\Delta T \to \delta t$, with the measurement time $\Delta t \to \infty$ allowed to be as large as necessary in parallel to eliminate misidentification errors.

With these assumptions, only double-bit-flips of the data qubits during the gate time $\delta t$ contribute to the logical error rate, with no delay between successive gate times. Thus,
\begin{align}
	\Gamma \tau &\approx 3\, \mu^2\tau \delta t = 12\, (\mu\tau)^2,
\end{align}
where we have taken the gate time to be $\delta t = 4 \tau$. 

The initial drop in fidelity of ancilla-based error correction comes from errors that would normally be detectable after multiple cycles, but do not have enough time to be detected before the final detection cycle ends. For the idealistic case, only data-qubit bit flips that happen after the initial CNOTs will cause an error, since bit flips before the CNOTs will be detected. Examining the gate timings in Fig.~\ref{fig:circuit} shows that only flips of bits 1 and 2 in the second half of the gate time can cause such an error, which yields a total fidelity drop of 
\begin{align}
    \Delta F_{\text{in}} = 2\mu(\delta t/2)=\mu\delta t = 4\mu\tau. 
\end{align}

\subsection{Logical error rate: Pessimistic case}\label{app:ancilla:pessimistic}

In the pessimistic case, logical errors can arise from several distinct mechanisms: (a) two data-qubit bit flips during the same cycle, (b) a single data-qubit bit flip during a CNOT gate, (c) one data-qubit bit flip and one ancilla-qubit bit flip, (d) two ancilla-qubit bit flips, (e) one misidentification and one bit flip, and (f) two misidentifications. For completeness, we also briefly consider a non-bit-flip error that is particularly problematic, (g) a single ancilla-qubit bit-phase-flip during a CNOT gate. Note that single ancilla-qubit bit flips are not problematic since they are simply corrected by the subsequent cycle. We now consider each logical error source in turn.

\begin{enumerate}[(a)]
 \item Two data-qubit bit flips 

This source of error is the same as in the idealistic case, but single bit flips during the gate time $\delta t$ are now dangerous.
In practice, however, $\Delta T \gg \delta t$ for the pessimistic case, so we can neglect all $\delta t/\Delta T$ corrections without harm. 
With this simplification, the two data-qubit bit-flip contribution to the logical error rate becomes $3\mu^2\Delta T$.

 \item One data-qubit bit flip during CNOT 

For specificity, we assume an initial 5-qubit state $|\psi\ra=(\alpha |000\ra+\beta|111\ra)|00\ra$, with an arbitrary logical superposition of data qubits 1,2, and 3, and the ancilla qubits 4 and 5 initialized in their ground states. We use the CNOT gate implementation outlined in Eq.~\eqref{eq:cnot}. Since the single-qubit gate corrections require only 1/4 the gate time, we assume that the relevant errors are caused by bit flips during the two-qubit gate and neglect single-qubit-gate errors. 
Since the $(Z_kX_j)_{90}$ gate is a continuous rotation that takes a finite time, a bit flip can occur at any intermediate angle of the rotation. 

An $(I_kX_j)$ bit flip of the ancilla commutes with this gate, so does not cause problems beyond those considered in subsequent sections. However, an $(X_kI_j)$ bit flip of a data qubit disrupts the gate rotation. When we combine the final single-qubit correction with the two-qubit gate disrupted at an angle $\phi\in[0,\pi/2]$, we find $(I_kX_j)_{-\pi/2}(Z_k X_j)_{\pi/2-\phi}(X_kI_j)(Z_k X_j)_\phi = (I_kX_j)_{-\pi/2}(Z_kX_j)_{\pi/2-2\phi}(X_kI_j) = X_{j|k}(4\phi-\pi)(X_kI_j) $, where 
\begin{align}
X_{j|k}(\theta) \equiv |0\ra\la 0|_k\otimes I_j + |1\ra\la 1|_k\otimes \exp(-i X_j\,\theta/2)
\end{align}
is a controlled-$X$ rotation of the ancilla. For clarity, we rescale the angle as $4\phi \equiv \varphi \in [0,2\pi]$.

Ignoring the phase corrections as unimportant here, we now consider each data-qubit bit-flip scenario of the complete gate sequence $(C_{5|3}C_{4|2}C_{5|2}C_{4|1})$ shown in Fig.~\ref{fig:circuit}:

\begin{itemize}
 \item $X_1$ during $C_{1|4}$:
 \begin{align*}
 &C_{5|3} C_{4|2}C_{5|2} (e^{i Z_1\pi/4}\otimes I_4)X_{4|1}(\varphi-\pi)(X_1I_4) |\psi\ra \nonumber \\
 &\qquad = (1-i)\sin(\varphi/2) (\alpha |100\ra-\beta|011\ra )|00\ra  \nonumber \\
 &\qquad + (i+1)\cos(\varphi/2) (\alpha |100\ra+\beta|011\ra)|10\ra 
 \end{align*}
 
 \item $X_2$ during $C_{5|2}$ 
 \begin{align*} 
 & C_{5|3} C_{4|2} (e^{i Z_2\pi/4}\otimes I_5)X_{5|2}(\varphi-\pi)(X_2I_5) C_{4|1}  |\psi\ra  \nonumber \\
 &\qquad = (1-i)\sin(\varphi/2) (\alpha |010\ra-\beta|101\ra )|10\ra \nonumber \\
 &\qquad + (i+1)\cos(\varphi/2) (\alpha |010\ra+\beta|101\ra )|11\ra
 \end{align*}
 
 \item  $X_2$ during $C_{4|2}$ 
 \begin{align*} 
 & C_{5|3} (e^{i Z_2\pi/4}\otimes I_4) X_{4|2}(\varphi-\pi)(X_2I_4) C_{5|2} C_{4|1}  |\psi\ra  \nonumber \\
 &\qquad = (1-i) \sin(\varphi/2) (\alpha |010\ra-\beta|101\ra )|00\ra \nonumber \\ 
 &\qquad + (i+1) \cos(\varphi/2) (\alpha |010\ra+\beta|101\ra )|10\ra
 \end{align*}
 
 \item $X_3$ during $C_{5|3}$ 
 \begin{align*} 
 & (e^{i Z_3\pi/4}\otimes I_5) X_{5|3}(\varphi-\pi)(X_3I_5) C_{4|2} C_{5|2} C_{4|1}  |\psi\ra  \nonumber \\
 &\qquad = (1-i)\sin(\varphi/2) (\alpha |001\ra-\beta|110\ra )|00\ra  \nonumber \\ 
 &\qquad + (i+1)\cos(\varphi/2) (\alpha |001\ra+\beta|110\ra)|01\ra 
 \end{align*}
\end{itemize}

After measuring the ancillas, each of these scenarios produces a logical phase flip with probability $\sin^2(\varphi/2)$. A phase flip can not be corrected by the bit flip code, so this is a logical error being induced by a bit flip. Averaging $\varphi$ over all possibilities yields a probability of $1/2$ for this to occur, per gate time of $\delta t/2$ in each cycle. Thus, the total phase-flip contribution to the logical error rate from all scenarios above is $4\,(1/2)\mu(\delta t/2)/\Delta T$.  

In the $C_{4|2}$ case, the ancilla measurement can additionally misidentify a bit-2 flip as a bit-1 flip with probability $\cos^2(\varphi/2)$; this produces a logical error since the subsequent correction cycle yields a syndrome of $(-1,-1)$ with an estimated encoding $XII$, which will be interpreted as a bit-3 flip to the encoding $XIX$ that is the inversion of the true $IXI$ encoding. Averaging over $\varphi$ again produces probability $1/2$, and a contribution to the logical error rate of $(1/2)\mu(\delta t/2)/\Delta T$.

The total contribution to the logical error rate from data-qubit bit-flips during a CNOT is thus $(5/4)\mu \delta t/\Delta T$. Since this contribution is linear in $\mu$, it is the dominant source of error for the ancilla-based approach. It is this term that is largely responsible for the smallness of the factor $\delta t/\Delta T$ upon optimization.

\item One data-qubit bit flip and one ancilla-qubit bit flip

A logical error can occur from flips within a single cycle when a data-qubit flip is combined with a misinterpretation of the syndrome measurement. Since data-qubit flips during the gate time have already been included in the logical error of the preceding subsections, we consider only flips that happen in the interval $[\delta t,\Delta T]$, with probability $\mu(\Delta T-\delta t)$. A syndrome misinterpretation can occur if an ancilla qubit is either incorrectly prepared, or flips just prior to readout. An incorrect preparation occurs if the ancilla flips during the second half of its heralding measurement $\Delta t/2$ before a detection cycle (see Fig.~\ref{fig:circuit}), since that flip will not be detected by the averaged record of the heralding measurement. Similarly, if the ancilla flips during the gate time $\delta t$ (assuming it commutes with the CNOT gates) or the first half of the final readout duration $\Delta t/2$, then the flip causes a misinterpretation of the syndrome. The total probability of these ancilla flips is thus $\mu(\Delta t + \delta t)$. There are four ways that combinations of flips can cause a logical error: Both 1+5 and 3+4 flips will misinterpret a bit 1 or bit 3 flip as sequence of two flips and cause a logical error in the next cycle. For example, 1+5 produces a true encoding of $XII$ but yields the two-cycle sequence of estimated encodings: $III\to IIX \to IXX$. Both 2+4 and 2+5 flips will misinterpret a bit 2 flip as a bit 1 or bit 3 flip and cause a logical error in the next cycle. For example, 2+4 produces a true encoding of $IXI$ but yields the two-cycle sequence $III\to XII \to XIX$. The total contribution to the logical error rate from a single cycle is therefore $4\,\mu^2(\Delta t + \delta t)(\Delta T - \delta t)/\Delta T$. 

A logical error can also occur from flips over two cycles when an ancilla qubit flips in one cycle and a data qubit flips in the next. In this case, the only situation that can cause a logical error is when the ancilla flips in the second half of the heralding measurement, with probability $\mu \Delta t/2$. This flip is unlikely to be detected in the current cycle, so will be postponed to the next cycle where it combines with the data-qubit flip to produce a logical error, with probability $\mu (\Delta T-\delta t)$. Tallying the same four ways this can produce a logical error produces the probability $2\mu^2\Delta t(\Delta T-\delta t)$ that this type of logical error occurs within a two cycle window. Given $N\gg 1$ total cycles, there are $N-1 \approx N$ windows of 2 consecutive cycles. The error rate contribution is therefore $N-1$ times the probability within a 2-cycle window over the total duration $N\Delta T$, which is $2 \mu^2 \Delta t (\Delta T-\delta t)/\Delta T$.
 
A similar situation arises when a data-qubit flips in one cycle followed by an ancilla-qubit flip in the first half of the syndrome measurement of the next cycle. The probability of obtaining a logical error from this scenario is similar to the previous two-cycle scenario, but includes the gate time. Therefore, the contribution of this type of error to the logical error rate is also $4 \mu^2 (\Delta t/2 + \delta t) (\Delta T-\delta t)/\Delta T$.
 
The total contribution of one data-qubit flip and one ancilla-qubit flip to the logical error rate is $8 \mu^2 \Delta t (\Delta T-\delta t)/\Delta T + 8\mu^2\delta t(\Delta T-\delta t)/\Delta T$.	Applying the condition $\delta t/\Delta T \ll 1$ yields the simplification $8\mu^2(\Delta t + \delta t)$.
 
 \item Two ancilla-qubit bit flips
 
A logical error can occur from flips in a single cycle when one ancilla flips during the first half of its syndrome measurement (producing an incorrect readout) and the remaining ancilla flips during the second half of its heralding measurement (producing an incorrect preparation for the next cycle). If the flips are ancilla 4 then ancilla 5, this produces the following sequence of estimated encodings over three cycles $III\to XII \to XXI \to XXX$. The reversed order of ancilla flips is also possible, so the contribution to the logical error rate is $2 \mu^2 (\Delta t/2)^2/\Delta T$.
 
A logical error can occur from flips over two cycles when the two ancillas sequentially flip during the first halves of their syndrome measurements, which produces the same result as above. Similarly, the same result occurs when the two ancillas sequentially flip during the second half of their heralding measurements. Thus, the logical error rate acquires two more contributions of $2 \mu^2 (\Delta t/2)^2/\Delta T$.

The total contribution of two ancilla-qubit flips to the logical error rate is thus $(3/2) \mu^2 \Delta t^2/\Delta T$.
 
 \item One misidentification and one bit flip
 
So far we have treated the ancilla measurements as ideal projective measurements, but in reality they are not. Since projective measurements implemented with dispersive coupling to microwave fields are thresholded continuous measurements---similar to those considered in the main text---the ancilla readout may be misidentified with probability $P_{\text{mis}}$ given in Eq.~\eqref{eq:pmis}. Such a misidentification error can cause a logical error when it is combined with a data-qubit flip or a flip in the other ancilla qubit. These contributions are much smaller than the preceding ones and can often be neglected; however, we include them for completeness.

The combinations of one misidentification and one bit flip follow the structure outlined in the preceding subsections that include at least one ancilla bit flip, so their contributions to the logical error rate can be listed here more compactly: A misidentification in the syndrome measurement of a cycle combined with a data-qubit flip in the same cycle contributes $4 P_{\text{mis}} \mu (\Delta T-\delta t)/\Delta T$. A misidentification in the heralding measurement of a cycle combined with a data-qubit flip in the same cycle contributes $4 P_{\text{mis}} \mu (\Delta T-\delta t)/\Delta T$. A misidentification in the heralding measurement of a cycle followed by a data-qubit flip in the following cycle contributes $4 P_{\text{mis}} \mu (\Delta T-\delta t)/\Delta T$. A misidentification in the syndrome measurement of a cycle combined with the complementary ancilla-qubit flip in the second half of its heralding measurement in the same cycle contributes $2P_{\text{mis}} \mu (\Delta t/2)/\Delta T$. A misidentification in the heralding measurement of a cycle combined with the complementary ancilla-qubit flip in the first half of its syndrome measurement in the same cycle contributes $2P_{\text{mis}} \mu (\Delta t/2)/\Delta T$. A misidentification in the heralding measurement of a cycle combined with the complementary ancilla-qubit flip in the second half of its heralding measurement of the following cycle contributes $2P_{\text{mis}} \mu (\Delta t/2)/\Delta T$. An ancilla-qubit flip in the first half of its syndrome measurement of a cycle followed by the complementary misidentification in the syndrome measurement of the following cycle contributes $2P_{\text{mis}} \mu (\Delta t/2)/\Delta T$. An ancilla-qubit flip in the second half of its heralding measurement of a cycle followed by the complementary misidentification in the heralding measurement of the following cycle contributes $2P_{\text{mis}} \mu (\Delta t/2)/\Delta T$.

The total contribution of one misidentification and one bit flip to the logical error rate is $12\,P_{\text{mis}}\,\mu(\Delta T - \delta t)/\Delta T + 5\,P_{\text{mis}}\,\mu\Delta t/\Delta T$. Applying the condition $\delta t / \Delta T \ll 1$ yields the simplification $12\,P_{\text{mis}}\,\mu + 5\,P_{\text{mis}}\,\mu\Delta t/\Delta T$.

 \item Two misidentifications 
 
Similarly to the case with two ancilla-qubit flips, two misidentifications can lead to a logical error. These errors are extremely small, but included for completeness. We list their contributions to the logical error rate compactly, since they follow the same structure as for two ancilla flips: A misidentification in the syndrome measurement of a cycle combined with a complementary misidentification in the heralding measurement of the same cycle contributes $2 P^2_{\text{mis}}/\Delta T$. A misidentification in the syndrome measurement of a cycle combined with the complementary misidentification in the syndrome measurement of the following cycle contributes $2 P^2_{\text{mis}}/\Delta T$. A misidentification in the heralding measurement of a cycle combined with the complementary misidentification in the heralding measurement of the following cycle contributes $2 P^2_{\text{mis}}/\Delta T$.

The total contribution of two misidentifications to the logical error rate is $6\,P^2_{\text{mis}}/\Delta T$. 

 \item One ancilla-qubit phase-bit flip during CNOT

For completeness, we highlight another type of logical error that could occur during a CNOT gate, but which we will ultimately neglect in the main text. In most of the analysis, we have been assuming pure bit flips $X$. However, when an environmental perturbation physically causes such a bit flip, the flip is likely to be a continuous rotation around an arbitrary axis of the Bloch sphere. As such, an ancilla-qubit bit flip may use an axis that does not commute with the rotation axis of the controlled-$X$ rotation in the CNOT. The worst case is a ``phase-bit'' flip of the ancilla, which uses the maximally non-commuting axis denoted by $I_kY_j$. 

Following the conventions of the preceding subsection, we find that the interrupted two-qubit gate sequence is $(I_kX_j)_{-\pi/2}(Z_kX_j)_{\pi/2-\phi}(I_kY_j)(Z_kX_j)_\phi = (I_kX_j)_{-\pi/2}(Z_kX_j)_{\pi/2-2\phi}(I_kY_j) = X_{j|k}(4\phi-\pi)(I_kY_j)$. After rescaling the angle as before $4\phi \equiv \varphi \in [0,2\pi]$, and noting that $Y_j = -iZ_jX_j$, we can immediately use the scenarios outlined in the preceding subsection to infer that an ancilla-phase-bit-flip during any of the four CNOT gate scenarios produces a logical phase-flip with probability $\cos^2(\varphi/2)$. Moreover, the $C_{5|2}$ case will produce a logical error with probability $\sin^2(\varphi/2)$ since a bit 2 flip is misidentified as a bit 1 flip. These cases would thus contribute another term of $(5/4)\mu\delta t/\Delta T$ to the logical error rate.

We also note in passing that a different implementation of a CNOT could result in a significantly worse error from such a phase-bit flip. Suppose that instead of the symmetric coupling of the $(Z_kX_j)_{90}$ gate, one directly implemented a controlled rotation gate $X_{j|k}(\pi)$. (The realistic gate operation is somewhere in between these idealizations.) A bit-phase flip in the middle of, e.g., the $C_{4|1}$ gate would then yield
\begin{align*} \nonumber
& C_{5|3} C_{4|2} C_{5|2} X_{4|1}(\pi-\phi)(I_1Y_4) X_{4|1}(\phi)|\psi\ra  \\ \nonumber
 &\qquad = (i-1)\sin(2\phi) \beta|111\ra |00\ra \\
 &\qquad +  (1+i)\big( \alpha |000\ra+\beta \cos(2\phi)|111\ra \big) |10\ra,
 \end{align*}
with similar outcomes for the other scenarios. In both cases of the ancilla measurement, the logical information is altered. In one case the logical state is completely projected and the information is destroyed, while in the other case the logical state is partially projected by a random amount. Neither of these types of logical error can be corrected by the three qubit code. 

In the spirit of analyzing how the bit-flip code protects against pure bit-flip errors, however, we do not include these sorts of errors caused by phase-bit-flips in our final estimate. This keeps the analysis focused solely on the performance of the code against the type of error for which it was intended.

 \end{enumerate}

Adding up all contributions to the logical error rate and neglecting $\delta t/\Delta T \ll 1$ corrections produces the total formula:
\begin{widetext}
\begin{align}
\Gamma\tau &= \frac{5}{4} \mu\tau \frac{\delta t}{\Delta T}+3 (\mu\tau)^2 \frac{\Delta T}{\tau} + 8 (\mu\tau)^2 \frac{\Delta t + \delta t}{\tau} + \frac{3}{2}(\mu\tau)^2 \frac{(\Delta t)^2}{ \tau\Delta T} + 12\, P_{\text{mis}}\,\mu\tau + 5\,P_{\text{mis}}\,\mu\tau \frac{\Delta t}{\Delta T} + 6\, P^2_{\text{mis}}\,\frac{\tau}{\Delta T}.
\end{align}
\end{widetext}

For both pessimistic and the optimistic methods, there are three contributions to the initial drop. First, if a data qubit flips after the CNOT gates, i.e. during the ancilla measurement time, then the flip is not detectable. Such a flip can occur with probability $3 \mu \Delta t$. Second, if an ancilla flips during the first half of its measurement time in the final cycle, then its reported parity outcome will be incorrect. Such an ancilla flip can occur with probability $2 \mu (\Delta t/2)$. Third, if an ancilla has its readout misidentified in the final cycle, then its reported parity outcome will be incorrect. Such a misidentification can occur with probability $2 P_{\text{mis}}$, with $P_{\text{mis}}$ as defined in Eq.~\eqref{eq:pmis}. Adding these contributions together, the drop in average fidelity from the final cycle is thus 
\begin{align}\label{eq:ancilladrop}
\Delta F_{\text{in}} &= 4 \mu\tau \frac{\Delta t}{\tau} + 2P_{\text{mis}}.
\end{align}

Using the optimization outlined in Eq.~\eqref{eq:maxt}, and methods similar to the main text, we can analytically obtain crude formulas for the optimized $\Delta F_{\text{in}}$ and $\Gamma$. We find that the optimized measurement time scales logarithmically with the bit-flip rate, 
\begin{align}
    \frac{\Delta t}{\tau} &\approx 2\ln\frac{3}{8\mu\tau\sqrt{\pi\ln[3/(8\mu\tau\sqrt{\pi\,\cdots})]}} \approx 2\ln\frac{1}{15\mu\tau},
\end{align}
where the final approximation applies in the range $\mu\tau\in[10^{-6},10^{-3}]$ by renormalizing the nested logarithmic dependence to a constant. The cycle time scales as the inverse square root, 
\begin{align}
    \frac{\Delta T}{\tau} &\approx \sqrt{\frac{5}{3\mu\tau}}.
\end{align}
These time scales produce an optimized logical error rate and initial drop of roughly
\begin{align}\label{eq:ancpessgamma}
\Gamma \tau &\approx 6.2(\mu\tau)^2[1 + 8\sqrt{\mu\tau}]\frac{\Delta T}{\tau} \approx 8\,(\mu\tau)^{3/2} \left[1 + 8 \sqrt{\mu \tau} \right], \nonumber \\
\Delta F_{\text{in}} &\approx 4\,(\mu\tau) \frac{\Delta t}{\tau} \approx 8\,(\mu\tau)\ln\frac{1}{15\mu\tau}. 
\end{align}
Due to the dangerous ancilla-flips during the CNOTs, the logical error rate dominantly scales as $\mu^{3/2}$ rather than $\mu^2$, which makes the pessimistic case significantly worse than the ideal case and unusable for practical error correction. Of all considered methods, it is the worst performing.

\subsection{Logical error rate: optimistic case}\label{app:ancilla:optimistic}

In the optimistic case, we assume that the 3-qubit code is a sub-code of a larger code that should be able to correct many of the errors introduced by the ancilla qubits in the pessimistic case. Most importantly, we assume that the phase-flip errors introduced by bit-flips during the CNOTs can be corrected by a suitable phase-flip encoding. We also assume that the code is fault-tolerant, so other errors that occur during the CNOT gates and ancilla measurements will be corrected. Similarly, we assume a non-Markovian  extensions to the 3-qubit code that can track the most likely past errors from observed sequences of syndrome measurements. 
With these enhancements in mind, the only important error will be the flip of two data-qubits, as intended by the 3-qubit code. This yields a logical error rate of $\Gamma = 3(\mu\tau)^2(\Delta T/\tau)$ that is quadratic in $\mu$ but linear in the cycle time $\Delta T$. For the initial drop in fidelity, the result Eq.~\eqref{eq:ancilladrop} in the pessimistic case also applies to the optimistic case.

Unlike the idealistic case, in order to reuse the ancilla qubits in consecutive cycles, the measurement time $\Delta t$ must be kept sufficiently long to ensure that misidentification errors remain rare, which bounds the cycle time from below $\Delta T \geq \delta t + \Delta t$ and limits how small one can make the logical error rate $\Gamma$. The operation of the 3-qubit code, even with extensions, assumes that flips occur rarely enough that $\mu\Delta T \ll 1$, so it is necessary that misidentifications also occur rarely, $P_{\text{mis}} \ll 1$.  A natural criterion for optimization is thus to keep the cycle time $\Delta T$ as small as possible while gracefully bounding $P_{\text{mis}}$ to be smaller than the bit flips being corrected, $P_{\text{mis}} \lesssim \mu\Delta t < \mu\Delta T \ll 1$, so that it scales appropriately with $\mu\tau$. 

Approximately solving the constraint $P_{\text{mis}} = \mu\Delta t$ produces a lower bound for the measurement time. In this small misidentification error regime, we can use the asymptotic formula $P_{\text{mis}} \approx \exp(-\Delta t/2\tau)/\sqrt{2\pi\Delta t/\tau}$ noted in Eq.~\eqref{eq:pmis}, which produces the consistency relation $\exp(-\Delta t/2\tau)/\sqrt{2\pi\Delta t/\tau} \lesssim \mu(\Delta t ) \ll 1$, yielding 
\begin{align}
  \frac{\Delta t}{\tau} &\gtrsim 
  2\ln\frac{1}{\mu\tau[2\ln(1/\mu\tau)]^{3/2}} \approx 2\ln\frac{1}{100\mu\tau}, \nonumber
\end{align}
assuming that $\Delta t$ will be minimized as the lower bound of the cycle time. The first approximation is valid in the range $\mu\tau\in[10^{-6},10^{-3}]$, while the second, more crude, approximation remains reasonably close over the same range. 

We thus obtain crude scaling formulas for the optimistic case of ancilla-based error correction:
\begin{align}\label{eq:ancoptgamma}
\Gamma \tau &\approx 3\,(\mu\tau)^2\frac{\Delta T}{\tau} \approx 3\,( \mu \tau)^2 \left[4+2\ln\frac{1}{100\mu \tau}\right], \\
\Delta F_{\text{in}} &\approx 6\,(\mu\tau)\frac{\Delta t}{\tau} \approx 12\,(\mu\tau)\ln\frac{1}{100\mu\tau}.
\end{align}
Importantly, for the optimistic case the scaling of $\Gamma$ with $\mu^2$ is restored, up to logarithmic corrections. However, the initial drop in fidelity remains significant because of bit flips in the extra two ancillary qubits during the final correction cycle. Of the analyzed ancilla-based cases, this optimistic case is the most plausible comparison to the continuous measurement filters in the main text. However, comparing the scaling must be done with care, since the timescale $\tau$ for ancilla-based measurement is likely a few times shorter than that of the parity measurements in the main text.

\end{document}